\documentclass{aa5}
\usepackage{graphics}
\begin{document}

%

   \title{A spectrophotometric atlas of Narrow-Line Seyfert 1 galaxies}
   \subtitle{}

   \author{M.-P. V\'eron-Cetty \inst{1}, P. V\'eron \inst{1} 
\and A.C. Gon\c{c}alves \inst{2}}

\offprints{P. V\'eron}

\institute{
Observatoire de Haute Provence, CNRS, F-04870 Saint-Michel l'Observatoire,
    France\\ 
\email{mira@obs-hp.fr; veron@obs-hp.fr}
\and
European Southern Observatory (ESO), Karl Schwarzschild Strasse 2, D-85748 
Garching bei M\"unchen, Germany\\
\email{adarbon@eso.org}}

   \date{Received ; accepted }

   \abstract{
  We have compiled a list of 83 objects classified as Narrow-Line Seyfert 1 
galaxies (NLS1s) or known to have a broad Balmer component narrower than 2\,000 
km s$^{-1}$. Out of these, 19 turned out to have been spectroscopically 
misidentified in previous studies; only 64 of the selected objects are genuine 
NLS1s. We have spectroscopically observed 59 of them and tried to characterize 
their Narrow and Broad-Line Regions (NLR and BLR) by fitting the emission-lines 
with Gaussian and/or Lorentzian profiles. \\
  In most cases, the broad Balmer components are well fitted by a single 
Lorentzian profile, confirming previous claims that Lorentzian rather than
Gaussian profiles are better suited to reproduce the shape of the NLS1s broad 
emission lines. This has consequences concerning their FWHMs and line ratios: 
when the broad Balmer components are fitted with a Lorentzian, most narrow line 
regions have line ratios typical of Seyfert 2s while, when a Gaussian profile is
used for fitting the broad Balmer components, the line ratios are widely 
scattered in the usual diagnostic diagrams (Veilleux \& Osterbrock 1987); 
moreover, the FWHM of the best fitting Lorentzian is systematically smaller than
the FWHM of the Gaussian.\\
 We find that, in general, the [O III] lines have a relatively narrow Gaussian 
profile ($\sim$ 200--500 km s$^{-1}$ FWHM) with often, in addition, a second 
broad ($\sim$ 500--1\,800 km s$^{-1}$ FWHM), blueshifted Gaussian component.
We do not confirm that the [O III] lines are weak in NLS1s. \\
 As previously suggested, there is a continuous transition of all properties 
between NLS1s and classical Broad-Line Seyfert 1 Galaxies (BLS1s) and the limit 
of 2\,000 km s$^{-1}$ used to separate the two species is arbitrary; R$_{4570}$,
the ratio of the Fe II to the H$\beta$ fluxes, could be a physically more 
meaningful parameter to distinguish them.
      \keywords{Galaxies -- Seyfert}
}

\titlerunning{Narrow-Line Seyfert 1 galaxies}
\authorrunning{V\'eron-Cetty et al.}
   \maketitle



\section{Introduction}

  Osterbrock \& Pogge (1985) have identified a class of AGNs having all  
properties of the Seyfert 1s with, however, very narrow Balmer lines and strong 
optical Fe II lines; they are called NLS1s. Quantitatively, a Seyfert 1 is 
called an NLS1 if the ``broad" component of the Balmer lines is narrower than 
2\,000 km s$^{-1}$ FWHM (Osterbrock 1987). NLS1s often have strong Fe II 
emission; many of them have a strong soft X-ray excess and display 
high-amplitude X-ray variability. \\

 The 2--10 keV spectrum of classical BLS1s can be fitted with a power-law with 
a photon index $\Gamma$=1.73$\pm$0.05 (Nandra \& Pounds 1994; Reynolds 1997; 
George et al. 1998a); in NLS1s, $\Gamma$ is significantly steeper ($\Gamma$=2.19
$\pm$0.10) (Leighly 1999b); in fact, it is anticorrelated with the H$\beta$ FWHM
(Brandt et al. 1997; Reeves \& Turner 2000). 

 Some Seyfert 1s show evidence for an excess of soft X-rays above the hard X-ray
power-law extrapolation, dominant below $\sim$1 keV (Saxton et al. 1993; George
et al. 2000). This excess is more important and more frequent in NLS1s than in 
BLS1s (Vaughan et al. 1999a; Leighly 1999b; Reeves \& Turner 2000). The soft
photon spectral index $\Gamma$ (0.1--2.4 keV), which measures the relative 
strength of the soft component, is correlated with the Balmer line width in the 
sense that steep soft X-ray spectra corresponds to narrow Balmer lines 
(Puchnarewicz et al. 1992; Boller et al. 1996; Wang et al. 1996; Laor et al. 
1997a). In fact NLS1s may show both steep as well as flat X-ray spectra, while 
BLS1s always have flat spectra (Grupe et al. 1999). However, several NLS1s show 
a significant intrinsic neutral hydrogen column density in excess of the 
Galactic value; the unability to detect a soft excess in the X-ray spectrum of 
some NLS1s could be due to the presence of such a high column density. Therefore
the soft excess may be more prevalent in NLS1s than observed (Leighly 1999b). 

 Most AGN spectra show the presence of a ``Big Blue Bump" (BBB) extending from 
optical frequencies upwards (Elvis et al. 1986; Sanders et al. 1989). The BBB 
has been interpreted as the thermal emission of a physically thin, optically 
thick accretion disk (Sun \& Malkan 1989; Siemiginowska et al. 1995).
 The soft X-ray spectral index is correlated to the strength of the ultraviolet 
bump in unabsorbed Seyfert 1s, indicating that the BBB is in fact an ultraviolet
to soft X-ray bump (Walter \& Fink 1993; Puchnarewicz et al. 1995; Page et al.
1999). NLS1s have significantly bluer spectra than BLS1s which is consistent 
with the presence of a more pronounced BBB in NLS1s (Grupe et al. 1998). 
However, when sufficient data are available, it seems that a single standard 
accretion disk model cannot fit the optical/UV/X-ray bump (Kolman et al. 1993; 
Wisotzki et al. 1995). \\

 NLS1s very frequently exhibit rapid and/or high-amplitude X-ray variability 
(Boller et al. 1996; Forster \& Halpern 1996; Molthagen et al. 1998). The 
X-ray-steep, narrow-H$\beta$ AGNs systematically show larger amplitude 
variations than the X-ray-flat, broad-H$\beta$ AGNs on time scales from 2 to 20
days (Fiore et al. 1998; Leighly 1999a; Turner et al. 1999b).
 Observed variability by a factor 2 in a few hours or less shows that a 
substantial fraction of the soft component comes from a compact region, smaller 
than a light-day. 
 Giant-amplitude X-ray variability (from one up to more than two orders of 
magnitude on a time scale of one to a few years) has been observed in several
NLS1s (see for instance Brandt et al. 1999 and Uttley et al. 1999). In the cases
of NGC~4051 and IRAS~13224$-$3809, changes by a factor of 10 or more have 
occured within a few hours (Leighly 1999a).
 It is quite remarkable that all these objects, except NGC~4051, have a very 
high soft photon index ($\Gamma>$4), at least when they are bright. 

 A few BLS1s have also displayed X-ray flux variability by a factor of 10 or 
more such as NGC~3227 ($\times$15) (Komossa \& Fink 1997a; George et al. 1998b)
and NGC~3786 ($\times$10) (Komossa \& Fink 1997b); in both cases, however, the 
variability has been attributed to a change in the column density of a warm 
absorber. \\

 A number of observations suggests that NLS1s are Seyfert 1s with a near- or 
super-Eddington accretion rate. 

 If the broad-line emitting region is gravitationally linked to the central 
black hole (BH), one can show that the FWHM of the lines depends on the mass of 
the BH, the ratio of the luminosity to the Eddington luminosity and the angle 
between the rotation axis of the gas disk and the line of sight. The NLS1s could
be either normal Seyfert 1s seen perpendicularly to the disk, or objects with a 
low mass BH radiating near the Eddington limit (Wang et al. 1996). 

 The combination of strong soft X-ray excess and steep power law prompted Pounds
et al. (1995) to postulate that NLS1s represent the supermassive BH analogue of
Galactic BH candidates (GBHC) in their high states. The high states of GBHCs are
thought to be triggered by increases in the accretion rate resulting in strong 
thermal emission from a disk accreting at the Eddington limit. 

 Standard accretion disks are not able to account for the soft X-ray excess 
unless the Eddington ratio is close to unity; a large accretion rate results in 
a more pronouced BBB which is shifted toward higher energies, resulting in 
stronger soft X-ray emission and hence steeper soft X-ray slope (Pounds et al. 
1987; Ross et al. 1992; Kuraszkiewicz et al. 2000). 

 At a fixed luminosity, BHs radiating at higher fractions of the Eddington rate 
will have lower masses; lower mass BHs are thought to be associated with 
physically smaller emission regions that vary more rapidly. This may explain 
why higher amplitude short term X-ray variability is observed in NLS1s (Fiore 
et al. 1998; Leighly 1999a; Turner et al. 1999b). 

 Nicastro (2000) proposed a model in which, for accretion rates \.M/\.M$_{\rm 
Edd}<$ 0.2 (sub-Eddington regime), the predicted FWHMs are quite broad 
($>$4\,000 km s$^{-1}$), while for \.M/\.M$_{\rm Edd}$=0.2-3 (Eddington to 
moderately super-Eddington), the corresponding FWHMs span the interval 
$\approx$1\,000-4\,000 km s$^{-1}$. \\

 The amount of published data on NLS1s increased dramatically over the last few 
years, specially since the launch of the {\it ROSAT} and {\it ASCA} satellites, 
and important progresses have been made in the X-ray domain. NLS1s 
however are still subject to much debate; our knowledge of the basic properties
concerning their emission-line regions (line profiles, line ratios, etc.), and 
of their relation to the X-ray properties, is still rather limited, as is the 
relationship between NLS1s and classical BLS1s. This is largely due to the fact
that little effort has been put in providing a set of high-quality optical 
spectroscopic data. Published data are very heterogeneous; spectra often have a
resolution insufficient to separate unambiguously the broad and narrow 
components of the Balmer lines; in addition, the presence of strong Fe II lines 
makes it difficult to measure H$\beta$. 

 Aware of the fact that a detailed and consistent study of NLS1 emission-line 
properties was missing and that the knowledge of these properties is of crucial
importance for understanding the basic physical differences between BLS1s and 
NLS1s and, ultimately, for fitting them into the standard unifying picture, we 
have obtained a homogeneous set of moderate resolution (3.4 \AA\ FWHM or 200
km s$^{-1}$ at H$\beta$) spectra around H$\alpha$ and/or H$\beta$ of a large 
number of NLS1s;   this setting turned out to be adequate as narrow line 
individual components are, in most cases, resolved with this resolution; 
however, in the few objects where a H II region is present near the galaxy 
nucleus a better resolution would of course allow to separate more easily the 
Seyfert 2 emission lines from the much narrower H II lines.

\section{Observations and data reduction}

  We have compiled all 83 objects known to us before January 1998 either to be 
NLS1s or to have a ``broad" Balmer component narrower than 2\,000 km s$^{-1}$, 
north of $\delta=-$25$^{\rm o}$, brighter than B=17.0 and with z $<$ 0.100. We 
have spectroscopically observed 76 of them. 

\begin{table}[h]
\caption{\label{observations}Observations dates and standard stars} 
\begin{flushleft}
\begin{tabular}{rcl}
\hline
Date~~~~~~~~~ & $\lambda$ range (\AA) & Standard stars \\
\hline
      21.03.95 & 6500 -- 7400 & BD~26\degr2606               \\
      31.08.95 & 4855 -- 5755 & Feige~15, BD~25\degr3941     \\
      01.09.95 & 4855 -- 5755 & \verb+  + and BD~28\degr4211 \\
      10.05.96 & 6700 -- 7600 & GD~140, BD~26\degr2606       \\
      11.05.96 & 4860 -- 5760 & Feige~98, Kopff~27           \\
15 -- 16.07.96 & 4675 -- 5575 & BD~28\degr4211               \\
      24.07.96 & 6335 -- 7235 & BD~28\degr4211               \\
      07.01.97 & 4720 -- 5620 & EG~247                       \\
09 -- 10.01.97 & 6175 -- 7075 & EG~247                       \\
04 -- 07.03.97 & 4825 -- 5725 & Feige~66                     \\
08 -- 12.03.97 & 6310 -- 7210 & Feige~66                     \\
      13.03.97 & 4825 -- 5725 & Feige~66                     \\
      29.10.97 & 6500 -- 6950 & Feige~24, EG~247             \\
      30.10.97 & 4825 -- 5280 & Feige~24, EG~247             \\
      31.10.97 & 6455 -- 7355 & Feige~24                     \\
01 -- 02.11.97 & 4655 -- 5555 & Feige~24                     \\
27 -- 29.05.98 & 4645 -- 5545 & Feige~66, Kopff~27           \\
      31.05.98 & 6430 -- 7330 & Feige~66                     \\
      15.06.98 & 4420 -- 6265 & GD~190                       \\
      16.06.98 & 6020 -- 7870 & GD~190                       \\
      23.09.00 & 4255 -- 6090 & Feige~15, EG~247             \\
 \hline
\end{tabular}
\begin{tabular}{lp{7.0cm}}
-- & On October 29 and 30, 1997, we have used a dispersion of 33 \AA\ mm$^{-1}$ 
      instead of 66 \AA\ mm$^{-1}$.\\
-- & In June 1998 and September 2000, we have used an EEV 42-20 instead of a 
TK 512 CCD \\
\end{tabular}
\end{flushleft}
\end{table}

 The observations were carried out during several observing runs with 
the spectrograph CARELEC (Lema\^{\i}tre et al. 1989) attached to the Cassegrain
focus of the Observatoire de Haute-Provence (OHP) 1.93 m telescope. Table
\ref{observations} gives the list of observing runs with the observed wavelength
ranges and the standard stars used. 

The detector was a 512$\times$512 pixels, 
27$\times$27 $\mu$m Tektronic CCD, except in June 1998 and September 2000 when 
we used a 1024$\times$2048 pixels, 13.5$\times$13.5 $\mu$m EEV 42-20 CCD. We 
generally used a 600 l\,mm$^{-1}$ grating resulting in a dispersion of 66 \AA\ 
mm$^{-1}$. On October 29 and 30, 1997, we used a dispersion of 33 \AA\ 
mm$^{-1}$. In each case, the galaxy nucleus was centered on the slit. Three to 
five columns of the CCD (3 to 5\arcsec) were extracted on the Tektronic CCD and 
7 column ($\sim$3\arcsec) on the EEV. The slit width was 2\farcs1, corresponding
to a projected slit width on the detector of 52 $\mu$m {\it i.e.} 1.9 and 3.8 
pixels with the Tektronic and the EEV CCD respectively. The resolution, as 
measured on the night sky lines, was $\sim$ 3.4 \AA\ FWHM. 

The spectra were flux calibrated using the standard 
stars given in Table \ref{observations}, taken from Oke (1974), Stone (1977), 
Oke \& Gunn (1983) and Massey et al. (1988).

 The spectra were analysed as described in V\'eron et al. (1980; 1981a, b) and 
Gon\c{c}alves et al. (1999a). Briefly, the three emission lines, H$\alpha$ and 
[N II]$\lambda\lambda$6548,6583 (or H$\beta$ and [O III]$\lambda\lambda$4959,
5007) were fitted by one or several sets of three Gaussian components; the width
and redshift of each component in a set were forced to be the same and the 
intensity ratios of the [N II] and [O III] lines were taken to be equal to their
theoretical values. The broad Balmer components were fitted by one or several 
Gaussian or Lorentzian profiles. 

 Nineteen galaxies turned out to have been misidentified as NLS1s; their spectra
will be published in V\'eron-Cetty et al. (2001). The 64 others are listed in 
Table \ref{catalog}. 

\begin{table*}
\caption{\label{catalog} List of NLS1s with z$<$0.100, B$<$17.0 and 
$\delta$ $>-$25$^{\rm o}$.
Col. 1: name, col. 2: short position, col. 3: the Galactic hydrogen column
density in units of 10$^{20}$ cm$^{-2}$ col. 4: redshift, col. 5: B magnitude, 
cols. 6 and 7: FWHM (in km s$^{-1}$) of the broad component of the Balmer lines, 
and reference, cols. 8 and 9: {\it ROSAT} (0.1--2.4 keV) photon index $\Gamma$ 
resulting from a power-law fit with Galactic absorption, and reference, col. 10:
{\it ROSAT} X-ray flux in cts s$^{-1}$ in the energy band 0.1--2.4 keV either from the
RASS catalogue or from the 1WGA catalogue (*), col. 11: 
``A" if we have obtained a red spectrum, ``B" for a blue spectrum.
 References: 
 (1)   Appenzeller \& Wagner 1991;
 (2)   Bade et al. 1995;
 (3)   Bassani et al. 1989;
 (4)   Boller et al. 1992;
 (5)   Boller et al. 1996;
 (6)   Boroson \& Green 1992;
 (7)   Boroson \& Meyers 1992;
 (8)   Brandt et al. 1995;
 (9)   Ciliegi \& Maccacaro 1996;
(10)   Goodrich 1989;
(11)   Green et al. 1989;
(12)   Grupe et al. 1998;
(13)   Grupe et al. 1999;
(14)   Leighly 1999b;
(15)   Maza \& Ruiz 1989;
(16)   Moran et al. 1996;
(17)   Netzer et al. 1987;
(18)   Osterbrock 1977a;
(19)   Osterbrock \& de Robertis 1985;
(20)   Osterbrock \& Pogge 1985;
(21)   Osterbrock \& Pogge 1987;
(22)   Osterbrock \& Shuder 1982;
(23)   Puchnarewicz et al. 1994;
(24)   Puchnarewicz et al. 1995;
(25)   Rush et al. 1996;
(26)   Stephens 1989;
(27)   Stirpe 1990;
(28)   Walter \& Fink 1993;
(29)   Wang et al. 1996;
(30)   Winkler 1992;
(31)   Yuan et al. 1998;
(32)   Zamorano et al. 1992;
(33)   Pfefferkorn et al. 2001.
}
\begin{center}
\begin{tabular}{llrllrrlrrl}
\hline
Name & Position & N$_{\rm H}$ & z & B & FW & & $\Gamma$ & & X \\
\hline

 Mark\,335          &0003$+$19 &  3.8 &0.025 & 13.7& 1640 &  (6) & 3.10  $\pm$0.05 & (14)&  2.48 &  AB \\
 I\,Zw\,1           &0050$+$12 &  5.1 &0.061 & 14.0& 1240 &  (6) & 3.09  $\pm$0.16 & (14)&  0.82 &  AB \\
 Ton\,S180          &0054$-$22 &  1.5 &0.062 & 14.4& 1000 & (30) & 3.04  $\pm$0.01 & (14)&  2.53 &  AB \\
 Mark\,359          &0124$+$18 &  4.8 &0.017 & 14.2&  480 & (22) & 2.4~~ $\pm$0.1  &  (5)&  0.61 &  AB \\
 MS\,01442$-$0055   &0144$-$00 &  2.8 &0.080 & 15.6& 1940 & (26) & 2.7~~ $\pm$0.2  &  (5)&  0.14 &  AB \\
 Mark\,1044         &0227$-$09 &  3.0 &0.016 & 14.3& 1280 & (10) & 3.08  $\pm$0.09 & (33)&  2.14 &  AB \\
 HS\,0328+0528      &0328$+$05 &  8.9 &0.046 & 16.7&      &      &                 &     &  0.22 &  AB \\
 IRAS\,03450$+$0055 &0345$+$00 & 11.1 &0.031 & 16.0& 1310 &  (7) &                 &     & *0.06 &~~-- \\
 IRAS\,04312$+$4008 &0431$+$40 & 34.5 &0.020 & 15.2&  690 & (16) & 2.8~~ $\pm$0.6  &  (4)&  0.16 &  AB \\
 Mark\,618          &0434$-$10 &  5.4 &0.036 & 14.5& 2300 &  (7) & 2.72  $\pm$0.15 & (25)&  0.58 &~~-- \\
 IRAS\,04416$+$1215 &0441$+$12 & 14.1 &0.089 & 16.1& 1670 & (16) & 2.96  $\pm$0.50 &  (4)&  0.16 &  AB \\
 IRAS\,04576$+$0912 &0457$+$09 & 13.5 &0.037 & 16.6& 1220 & (16) &                 &  (4)&       &  AB \\
 IRAS\,04596$-$2257 &0459$-$22 &  3.1 &0.041 & 15.6& 1500 & (11) &                 &     & *0.51 &~~-- \\
 IRAS\,05262$+$4432 &0526$+$44 & 38.3 &0.032 & 13.6&  700 & (16) &                 &     &  0.06 &  AB \\
 RX\,J07527$+$2617  &0749$+$26 &  5.1 &0.082 & 17.0& 1000 &  (2) & 3.00  $\pm$0.26 &  (2)&  0.16 &  AB \\
 Mark\,382          &0752$+$39 &  5.8 &0.034 & 15.5& 1500 & (18) & 3.09  $\pm$0.23 & (33)&  0.45 &  AB \\
 Mark\,110          &0921$+$52 &  1.6 &0.036 & 15.4& 2120 &  (6) & 2.35  $\pm$0.05 & (29)&  1.69 &~~-- \\
 Mark\,705          &0923$+$12 &  4.0 &0.028 & 14.9& 1990 &  (6) & 2.33  $\pm$0.09 & (33)&  1.25 &  AB \\
 Mark\,707          &0934$+$01 &  4.7 &0.051 & 16.3& 1320 &  (6) & 2.40            & (29)&  0.46 &  AB \\
 Mark\,124          &0945$+$50 &  1.3 &0.056 & 15.3& 1400 & (18) &                 &     &       &  AB \\
 Mark\,1239         &0949$-$01 &  4.1 &0.019 & 14.4&  910 & (20) & 2.94  $\pm$0.14 & (25)&  0.05 &  AB \\
 IRAS\,09571$+$8435 &0957$+$84 &  3.9 &0.092 & 17.0& 1120 & (16) & 1.39  $\pm$0.40 &  (4)&  0.07 &  AB \\
 PG\,1011$-$040     &1011$-$04 &  4.5 &0.058 & 15.5& 1440 &  (6) &                 &     &       &  AB \\
 PG\,1016$+$336     &1016$+$33 &  1.6 &0.024 & 15.9& 1600 & (21) &                 &     &       &  AB \\
 Mark\,142          &1022$+$51 &  1.2 &0.045 & 15.8& 1620 &  (6) & 3.15  $\pm$0.11 & (14)&  1.75 &  AB \\
 KUG\,1031$+$398    &1031$+$39 &  1.4 &0.042 & 15.6& 1500 & (24) & 4.15  $\pm$0.10 & (14)&  2.66 &  AB \\
 RX\,J10407$+$3300  &1037$+$33 &  2.2 &0.081 & 16.5& 1700 &  (2) & 2.13  $\pm$0.15 &  (2)&  0.25 &  AB \\
 Mark\,734          &1119$+$12 &  2.7 &0.049 & 14.6& 1820 &  (6) & 3.63  $\pm$0.19 & (33)&  0.42 &  AB \\
 Mark\,739E         &1133$+$21 &  2.2 &0.030 & 14.1&  900 & (17) & 2.43  $\pm$0.14 & (33)&  0.49 &  AB \\
 MCG\,06.26.012     &1136$+$34 &  1.9 &0.032 & 15.4& 1685 & (13) & 2.77  $\pm$0.08 &  (9)&  0.86 &  AB \\
 Mark\,42           &1151$+$46 &  1.9 &0.024 & 15.4&  670 & (20) & 2.76  $\pm$0.23 & (33)&  0.19 &  AB \\
 NGC\,4051          &1200$+$44 &  1.3 &0.002 & 12.9&  990 & (22) & 2.84  $\pm$0.04 & (14)&  3.92 &  AB \\
 PG\,1211$+$143     &1211$+$14 &  2.8 &0.085 & 14.6& 1860 &  (6) & 3.03  $\pm$0.15 & (14)&  1.56 &  AB \\
 Mark\,766          &1215$+$30 &  1.8 &0.012 & 13.6& 2400 & (20) & 2.79  $\pm$0.11 & (14)&  4.71 &  AB \\
 MS\,12170$+$0700   &1216$+$07 &  2.2 &0.080 & 16.3&      &      &                 &     &       &  AB \\
 MS\,12235$+$2522   &1223$+$25 &  1.8 &0.067 & 16.3& 1730 & (26) & 3.9~~ $\pm$0.3  &  (5)& *0.52 &  AB \\
 IC\,3599           &1235$+$26 &  1.4 &0.021 & 15.6& 1200 &  (8) & 4.2~~ $\pm$0.1  & (12)&  5.10 &  AB \\
 PG\,1244$+$026     &1244$+$02 &  1.9 &0.048 & 16.1&  830 &  (6) & 3.26  $\pm$0.13 & (14)&  1.30 &  AB \\
 NGC\,4748          &1249$-$13 &  3.6 &0.014 & 14.0& 1100 & (19) & 2.46  $\pm$0.15 & (26)&  0.97 &  AB \\
 Mark\,783          &1300$+$16 &  2.0 &0.067 & 15.6& 1900 & (20) &                 &     &  0.29 &  AB \\
 R\,14.01           &1338$-$14 &  7.6 &0.042 & 14.6& 1790 & (15) &                 &     &  0.31 &  AB \\
 Mark\,69           &1343$+$29 &  1.1 &0.076 & 15.9& 1500 & (18) &                 &     &  0.26 &  AB \\
 2E\,1346$+$2646    &1346$+$26 &  1.1 &0.059 & 16.5&      &      & 2.68  $\pm$0.2  & (23)& *0.16 &  AB \\
 PG\,1404$+$226     &1404$+$22 &  2.0 &0.098 & 15.8&  880 &  (6) & 4.04  $\pm$0.20 & (29)&  0.45 &  AB \\
 Mark\,684          &1428$+$28 &  1.5 &0.046 & 14.7& 1400 & (21) & 2.4~~ $\pm$0.2  & (12)&  0.58 &  AB \\
 Mark\,478          &1440$+$35 &  1.0 &0.077 & 14.6& 1450 &  (6) & 3.06  $\pm$0.03 & (14)&  5.78 &  AB \\
 PG\,1448$+$273     &1448$+$27 &  2.7 &0.065 & 15.0&  910 &  (6) & 3.17  $\pm$0.32 & (28)&  0.78 &  AB \\

\hline
\end{tabular}
\end{center}
\end{table*}
\addtocounter{table}{-1}
\begin{table*}
\caption{(end)}
\begin{center}
\begin{tabular}{llrllrrlrrl}
\hline
Name & Position & N$_{\rm H}$ & z & B & FW & & $\Gamma$ & & X \\
\hline

 IRAS\,15091$-$2107 &1509$-$21 &  8.8 &0.044 & 14.8& 1480 & (10) &                 &     &  0.36 &~~-- \\
 MS\,15198$-$0633   &1519$-$06 & 12.4 &0.084 & 14.9& 1304 &  (1) & 3.39 $\pm$0.26  & (31)&  0.15 &  A  \\
 Mark\,486          &1535$+$54 &  1.8 &0.038 & 14.8& 1480 &  (6) &                 &     & *0.05 &  AB \\
 IRAS\,15462$-$0450 &1546$-$04 & 12.5 &0.100 & 16.4&      &      &                 &     &       &  AB \\
 Mark\,493          &1557$+$35 &  2.0 &0.031 & 15.1&  410 & (20) & 2.84  $\pm$0.14 & (33)&  0.52 &  AB \\
 EXO\,16524$+$3930  &1652$+$39 &  1.7 &0.069 & 16.7& 1000 &  (3) & 2.7~~ $\pm$0.2  &  (5)&  0.10 &  AB \\
 B3\,1702$+$457     &1702$+$45 &  2.2 &0.060 & 15.1&  490 & (16) & 2.37  $\pm$0.18 & (14)&  0.91 &  AB \\
 RX\,J17450$+$4802  &1743$+$48 &  3.1 &0.054 & 15.9& 1600 &  (2) & 2.64  $\pm$0.13 &  (2)&   0.27&~~$\;$B \\
 Kaz\,163           &1747$+$68 &  4.4 &0.063 & 15.0& 1260 & (10) & 2.76  $\pm$0.03 & (14)&  0.21 &  AB \\
 Mark\,507          &1748$+$68 &  4.3 &0.053 & 15.4&  965 & (10) & 1.68  $\pm$0.16 & (14)& *0.03 &  AB \\
 HS\,1817$+$5342    &1817$+$53 &  4.9 &0.080 & 15.2&      &      &                 &     &  0.48 &  AB \\
 HS\,1831$+$5338    &1831$+$53 &  4.9 &0.039 & 15.9&      &      &                 &     &  0.06 &  AB \\
 Mark\,896          &2043$-$02 &  4.0 &0.027 & 14.6& 1330 & (27) & 3.38  $\pm$0.05 & (33)&  0.44 &  AB \\
 MS\,22102$+$1827   &2210$+$18 &  6.2 &0.079 & 16.7&      &      &                 &     &  0.13 &  AB \\
 Akn\,564           &2240$+$29 &  6.4 &0.025 & 14.2&  750 & (27) & 3.47  $\pm$0.07 & (14)&  3.84 &  AB \\
 HS 2247$+$1044     &2247$+$10 &  6.2 &0.083 & 15.8&      &      &                 &     &  0.18 &  AB \\
 Kaz 320            &2257$+$24 &  4.9 &0.034 & 16.8& 1800 & (32) &                 &     &  0.53 &  AB \\

\hline
\end{tabular}
\end{center}
\end{table*}
\normalsize

 Before analysing our blue spectra, the Fe II multiplets were removed
following the method described in Boroson \& Green (1992). This consists of 
subtracting a suitable fraction of a Fe II template from the NLS1 spectrum so 
that the flux and width of the H$\beta$ and [O III] lines are no longer affected
by the underlying multiplet emission. Such a template is usually obtained by 
taking a high signal-to-noise spectrum of I~Zw~1, an NLS1 showing strong narrow 
Fe II emission, from which the H$\beta$ and [O III] lines are carefully removed.
We have observed I~Zw~1 with the same instrumental setting as the rest of the 
galaxies in our sample and used it to build an Fe II template, following this 
method. 

\begin{figure*}
\resizebox{!}{20.2cm}{\includegraphics{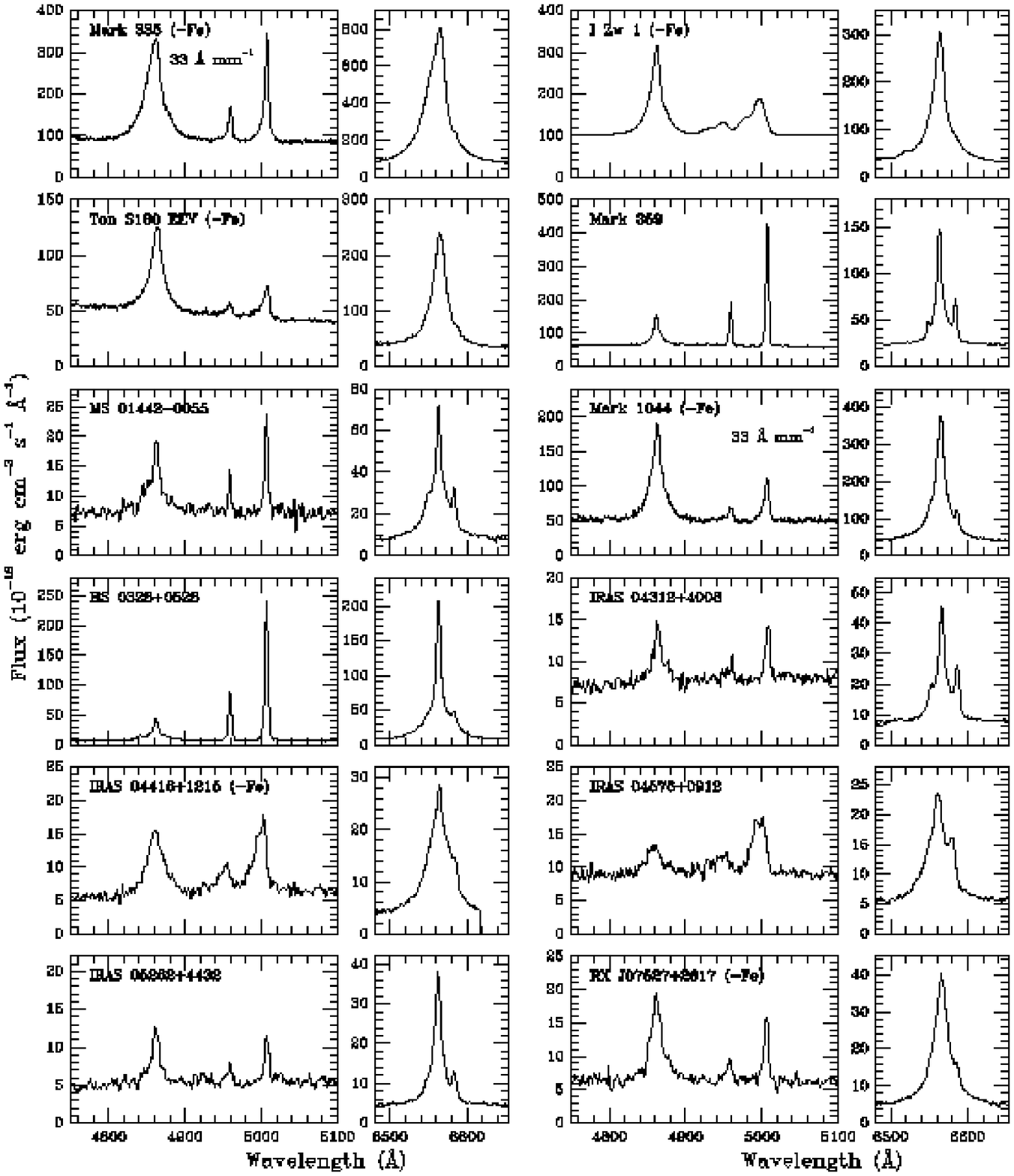}}
\caption{\label{nls1_1}Deredshifted blue and red spectra of the observed NLS1s.
}
\end{figure*}

\begin{figure*}
\resizebox{!}{20.2cm}{\includegraphics{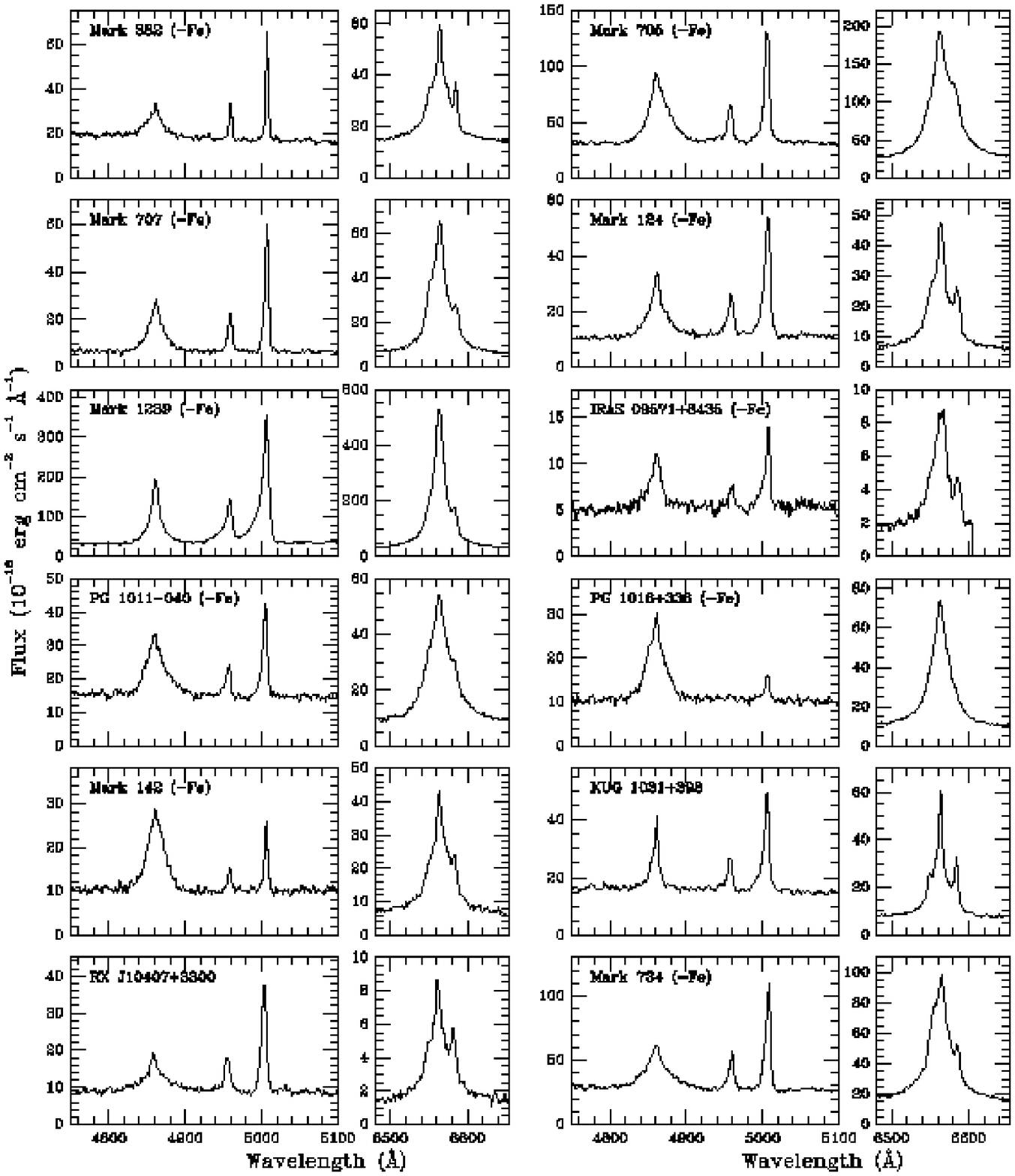}}
\caption{\label{nls1_2}Deredshifted blue and red spectra of the observed NLS1s (continued).
}
\end{figure*}

\begin{figure*}
\resizebox{!}{20.2cm}{\includegraphics{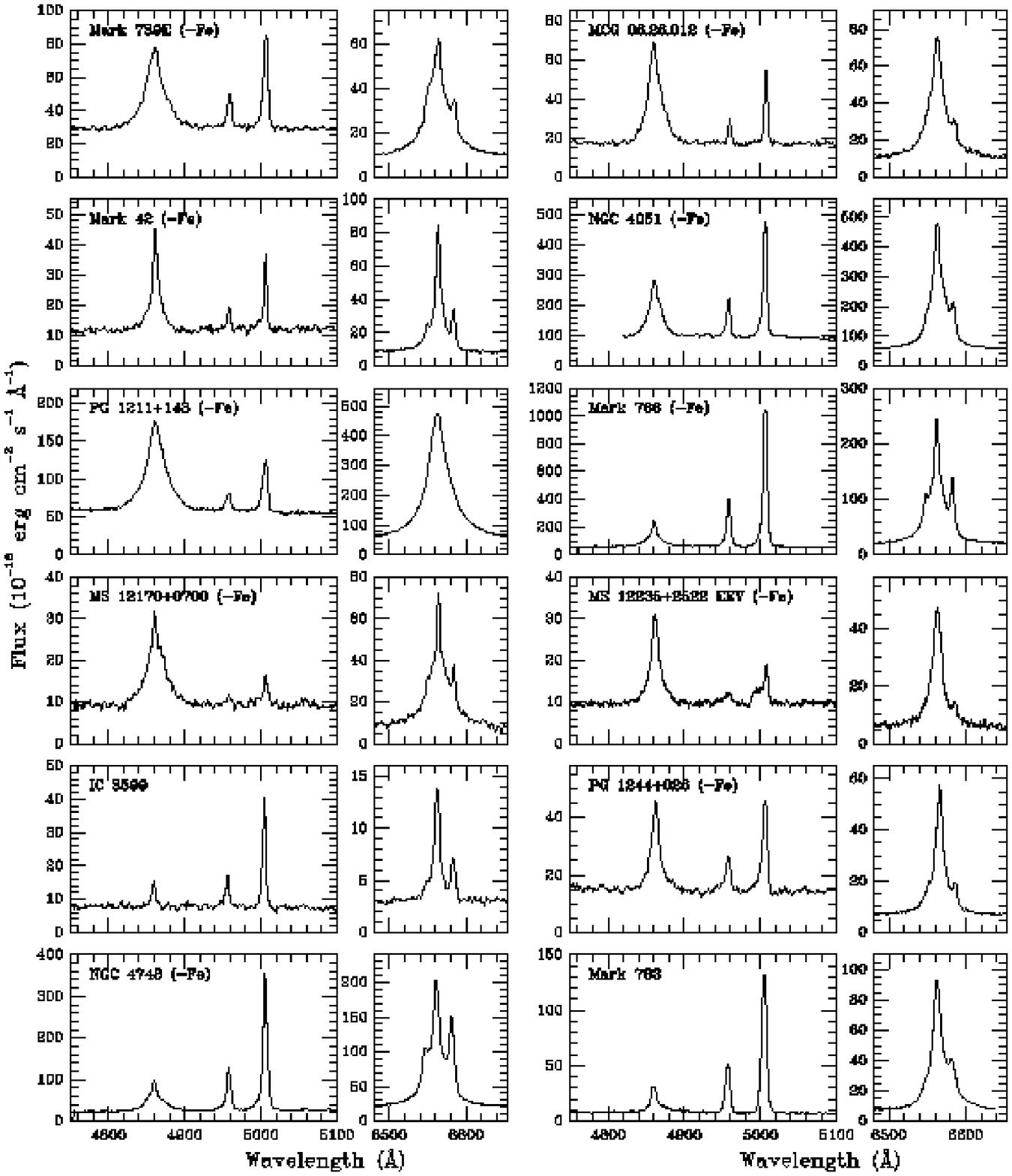}}
\caption{\label{nls1_3}Deredshifted blue and red spectra of the observed NLS1s (continued).
}
\end{figure*}

\begin{figure*}
\resizebox{!}{20.2cm}{\includegraphics{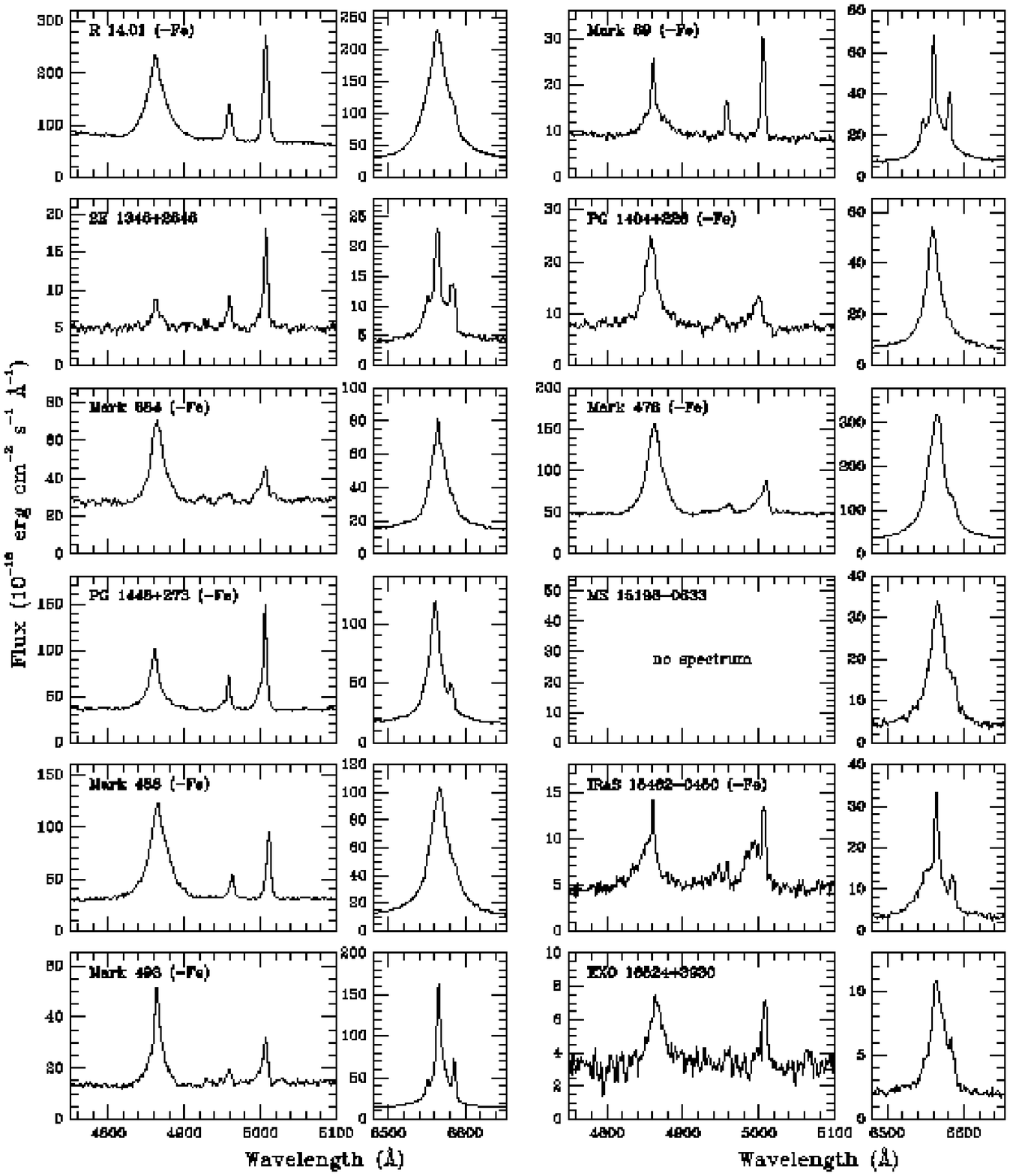}}
\caption{\label{nls1_4}Deredshifted blue and red spectra of the observed NLS1s (continued).
}
\end{figure*}

\begin{figure*}
\resizebox{!}{20.2cm}{\includegraphics{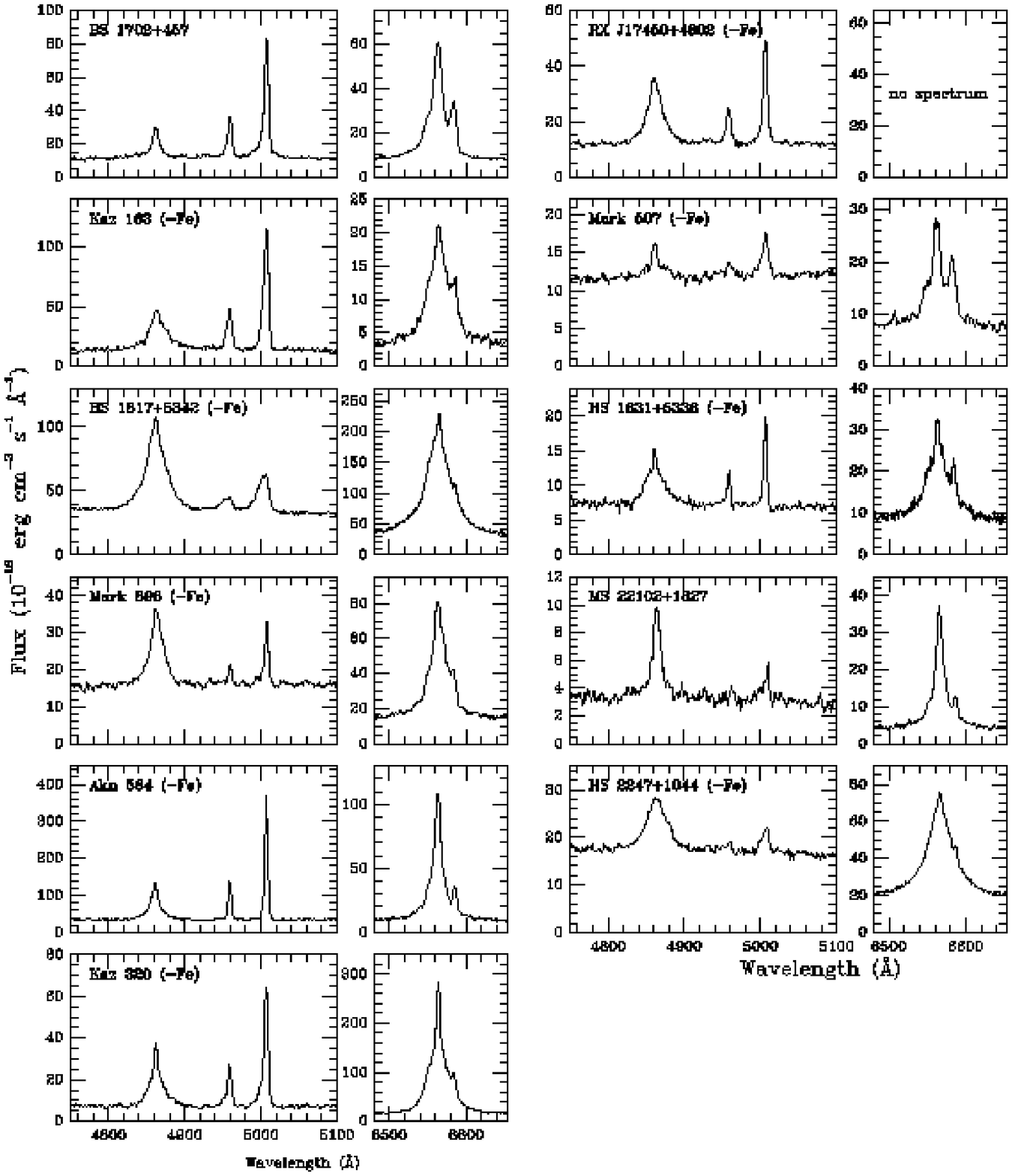}}
\caption{\label{nls1_5}Deredshifted blue and red spectra of the observed NLS1s (end).
}
\end{figure*}

Our spectra are shown in Figs. \ref{nls1_1} to \ref{nls1_5}.

 The Fe II strength is usually quantified by R$_{4570}$ = Fe II 
$\lambda$4570/H$\beta$, {\it i.e.} the ratio of the fluxes of the $\lambda$4570 
\AA\ blend measured between $\lambda$4434 and $\lambda$4684 and of H$\beta$,
including the narrow component (Boroson \& Green 1992). 

\begin{figure}[t]
\resizebox{8.8cm}{!}{\includegraphics{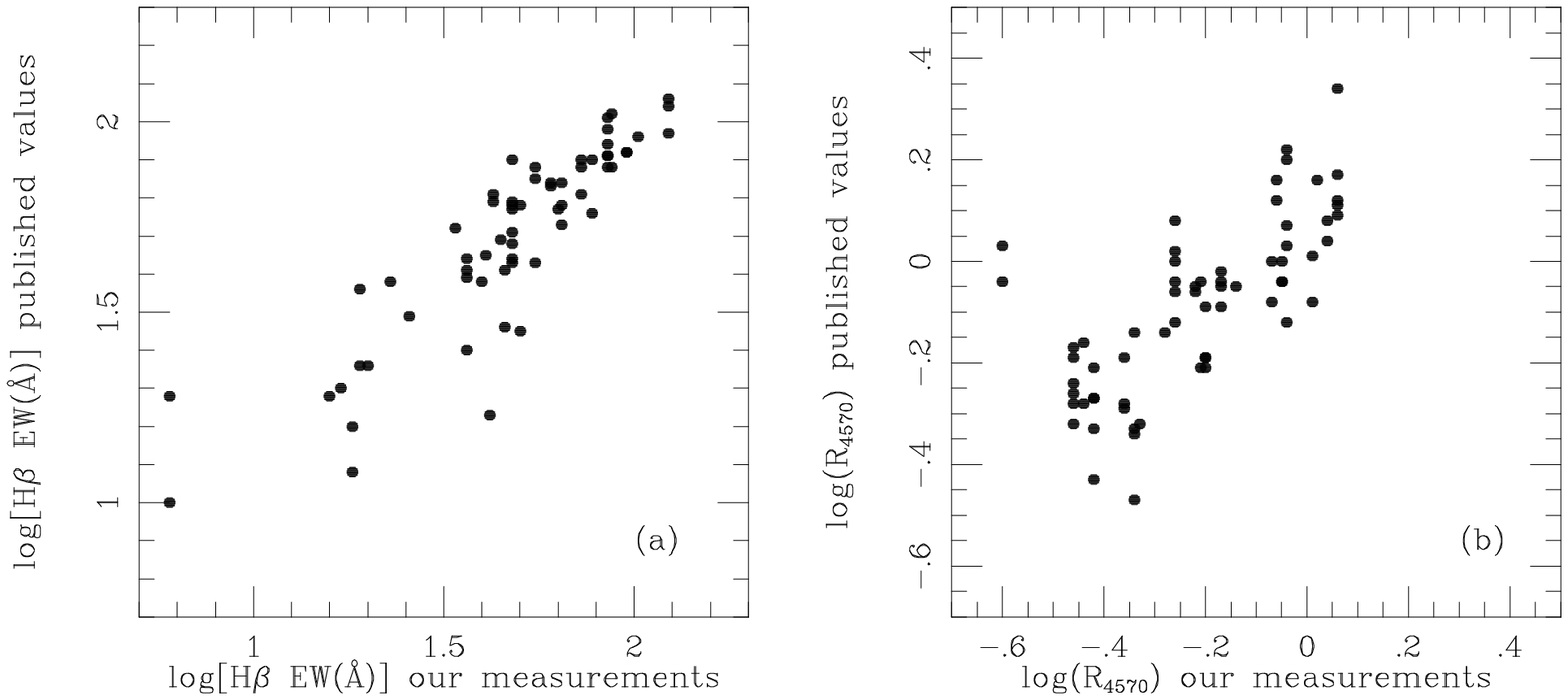}}
\caption{\label{ewhb}
(a) Comparison between our measurements of the H$\beta$ EW and the published 
ones. (b) Comparison between our measurements of R$_{4570}$ and the published 
ones. 
}
\end{figure}

 We have measured the H$\beta$ equivalent width (EW) and the ratio R$_{4570}$ 
for all objects in our sample. Figure \ref{ewhb} shows plots of our measurements 
{\it vs} published ones for the H$\beta$ EW (a) and R$_{4570}$ (b). 
Our values of the H$\beta$ EW are in good agreement with those of Boroson \& 
Green (1992); the ratios for the 13 objects in common have a mean value of 1.0 
with a dispersion of 12\%. Our values are also in good agreement with those of 
Goodrich (1989) with however a larger dispersion (40\%); this could be due 
partly to variability. Our measurements of R$_{4570}$ are 25\% lower than the 
values published by Boroson \& Green (1992) (13 objects in common) and 30\% 
lower than those of Goodrich (1989) (10 objects); in both cases the dispersion 
is $\sim$25\%. A significant difference exists for NGC~4051. 

\section{Results and discussion}

\subsection{The broad line region}

\subsubsection{High- and low-ionization lines}

 In Seyfert 1s, the broad emission lines can be separated into two distinct 
systems: the ``high-ionization lines" (HILs): C IV $\lambda$1550, He II 
$\lambda$4686, He II $\lambda$1640, etc, and the ``low-ionization lines" (LILs):
Fe II, Mg II $\lambda$2800, etc. (Collin-Souffrin \& Lasota 1988; Gaskell 2000).

 The HILs and LILs show strong kinematic differences (Sulentic et al. 1995).
C~IV is systematically broader than Mg~II (Mathews \& Wampler 1985) or H$\beta$
(Wang et al. 1998) and shows a strongly blueshifted and blue asymmetric profile 
(Marziani et al. 1996), and He~II $\lambda$4686 systematically broader than 
H$\beta$ (Boroson \& Green 1992; Peterson et al. 2000).  The widths of 
H$\beta$ and Fe~II (Boroson \& Green 1992) and of H$\beta$ and 
C~III]$\lambda$1909 (Wills et al. 2000) are strongly correlated ; in two NLS1s,
Leighly (2001) found that Si~III] and C~III] have the same width as the LILs;
but, if Puchnarewicz et al. (1997) found that the FWHM of H$\beta$ and Mg~II are
also correlated, they also found that that of C~III] and Mg~II are not. These 
observations support the idea of separate HIL and LIL emitting regions although 
it is not quite clear if C~III] belongs to the HIL or to the LIL region. \\

 The Fe II emission in most AGNs is too strong to be explained by 
photoionization (Phillips 1978b; Kwan et al. 1995); Fe II lines, and the other 
LILs as well, are more likely to 
be explained in the framework of collisional models where both the excitation of
the lines and the ionization of the elements are due to collisions in a high 
density optically-thick medium illuminated mainly by hard X-rays; the Fe II 
emission region has a high density (10$^{10}$ $<$ N$_{\rm e}$ $<$ 10$^{11}$ or 
10$^{12}$ cm$^{-3}$), a high column density (N$_{\rm H}>$ 10$^{24}$ cm$^{-2}$) 
and a low temperature (7\,500 $<$ T $<$ 10\,000 K) (Collin-Souffrin et al. 1980;
1988b; Joly 1981; Clavel et al. 1983; Collin-Souffrin \& Lasota 1988; Kwan et 
al. 1995).  There is a significant correlation between 
Si~III]$\lambda$1892/C~III]$\lambda$1909 and R$_{4570}$; Si~III and C~III have 
similar ionization potentials, but Si~III] has a critical density more than 
one order of magnitude larger than C~III] (1.1$\times$10$^{11}$ and 5$\times$10$^{9}$ 
cm$^{-3}$ respectively); Si~III]/C~III] is thus a density indicator and becomes
larger when density is higher; Si~III]/C~III] increases with increasing 
N$_{\rm e}$
up to N$_{\rm e}$=10$^{11}$ cm$^{-3}$ (Aoki \& Yoshida 1999; Wills et al. 1999; 
Kuraszkiewicz et al. 2000). It seems therefore likely that the C III] line comes
from the same high density region as H$\beta$ and Fe II where it is at least 
partially suppressed by collisions.

 HILs are emitted by low-pressure optically thin clouds (density of at most a 
few 10$^{9}$ cm$^{-3}$) illuminated by a rather soft continuum radiation (UV and
soft X-rays) (Collin-Souffrin \& Lasota 1988). Collin-Souffrin et al. (1988a) 
suggested that these lines are produced in clouds undergoing predominantly 
outward motions along the system axis, the clouds receding from us being hidden 
by an opaque structure such as the accretion disk. 
 The bipolar outflow could be a hydromagnetically driven wind accelerated 
radiatively and centrifugally away from the surface of the accretion disk 
(Emmering et al. 1992; K\"onigl \& Kartje 1994; Bottorf et al. 1997; Murray \& 
Chiang 1998). \\

 The broad Balmer lines exhibit a wide variety of profile shapes and a large 
range in width (Osterbrock \& Shuder 1982; de Robertis 1985; Crenshaw 1986; 
Stirpe 1991; Miller et al. 1992); they are often strongly asymmetric (Corbin 
1995). The BLR consists of two 
components: one Intermediate Line Region (ILR) with line width $\sim$2\,000 km 
s$^{-1}$ FWHM, with the peak within a few hundred kilometers per second of the 
systemic redshift, and a Very Broad Line Region (VBLR) with lines of width 
$>$ 7\,000 km s$^{-1}$ and blueshifted by more than 1\,000 km s$^{-1}$; 
differences in the relative strengths of these components account for much of 
the diversity of broad line profiles (Wills et al. 1993; Brotherton et al. 1994;
Corbin 1995; 1997; Francis et al. 1992). The spectra of the VBLR and ILR are 
very different (Brotherton et al. 1994); the VBLR and the ILR can probably be 
identified with the HIL and LIL regions respectively (Puchnarewicz et al. 1997).
This is confirmed by variability studies: the profile of the broad emission 
lines are variable; many of them can be 
described by two Gaussian components that are nearly stationary in wavelength, 
and which vary independently of one another in relative flux (Peterson et al. 
1999). This is the case for NGC~5548 (Dumont et al. 1998) and NGC~3516 (Goad 
et al. 1999) in which the emission lines are best explained by the superposition
of an emission line cloud with variable lines and another which shows no 
variability; the emission spectrum of the non variable cloud is dominated by 
Balmer lines and Fe II emission. In the case of PG~1416$-$129, the broad
H$\beta$ component (4\,000 km s$^{-1}$ FWHM) is strongly variable while the very
broad component (13\,000 km s$^{-1}$ FWHM) has a much smaller amplitude 
(Sulentic et al. 2000b).  \\

 LILs could be produced in the outer part of the disk itself as a result of 
energy reflected from the flow above the disk (Collin-Souffrin et al. 1988a). 
The profile of the lines produced in a disk are generally not double peaked, 
except if the radius of the disk is small ($\sim$10$^{3}$ R$_{\rm G}$ where 
R$_{\rm G}$ is the gravitational radius of the central BH), in which case the 
line intensities are small, the line profiles are U-shaped and very broad; if 
the disk radius is large ($>$10$^{4}$ R$_{\rm G}$), the line intensities are 
large and their profile generally single peaked;  the spectrum consists mainly of
LILs and can be a major part of the broad line emission (Dumont \& 
Collin-Souffrin 1990); these lines are, under certain conditions, more similar 
to a Lorentzian than to a Gaussian (see their Fig. 4b).

\subsubsection{On the profile of the NLS1 broad lines}

  According to Moran et al. (1996) and Leighly (1999b), many NLS1s have 
symmetric emission lines with more nearly Lorentzian than Gaussian profiles.
However Rodr\'{\i}guez-Ardila et al. (2000b) claimed that Lorentzian profiles
are not suited to represent the NLS1 broad emission-lines; this conclusion 
is based on the fact that these authors were unable to get a good fit when using
a single Lorentzian for each of the emission lines; however, there is a major 
inconsistency in their procedure: they assumed that the Balmer lines (either 
H$\alpha$ or H$\beta$) had a pure Lorentzian profile, not allowing for the 
presence of a narrow component.
 It so happens that one of the objects they present as an example of the poor 
results obtained when fitting the emission lines with Lorentzians (Mark~1239) 
is one of those for which we obtain a very good fit using a Lorentzian profile 
for the broad Balmer line components in addition to a narrow component having 
the same Gaussian profile as the forbidden lines. 

\begin{figure*}[t]
\resizebox{17.5cm}{!}{\includegraphics{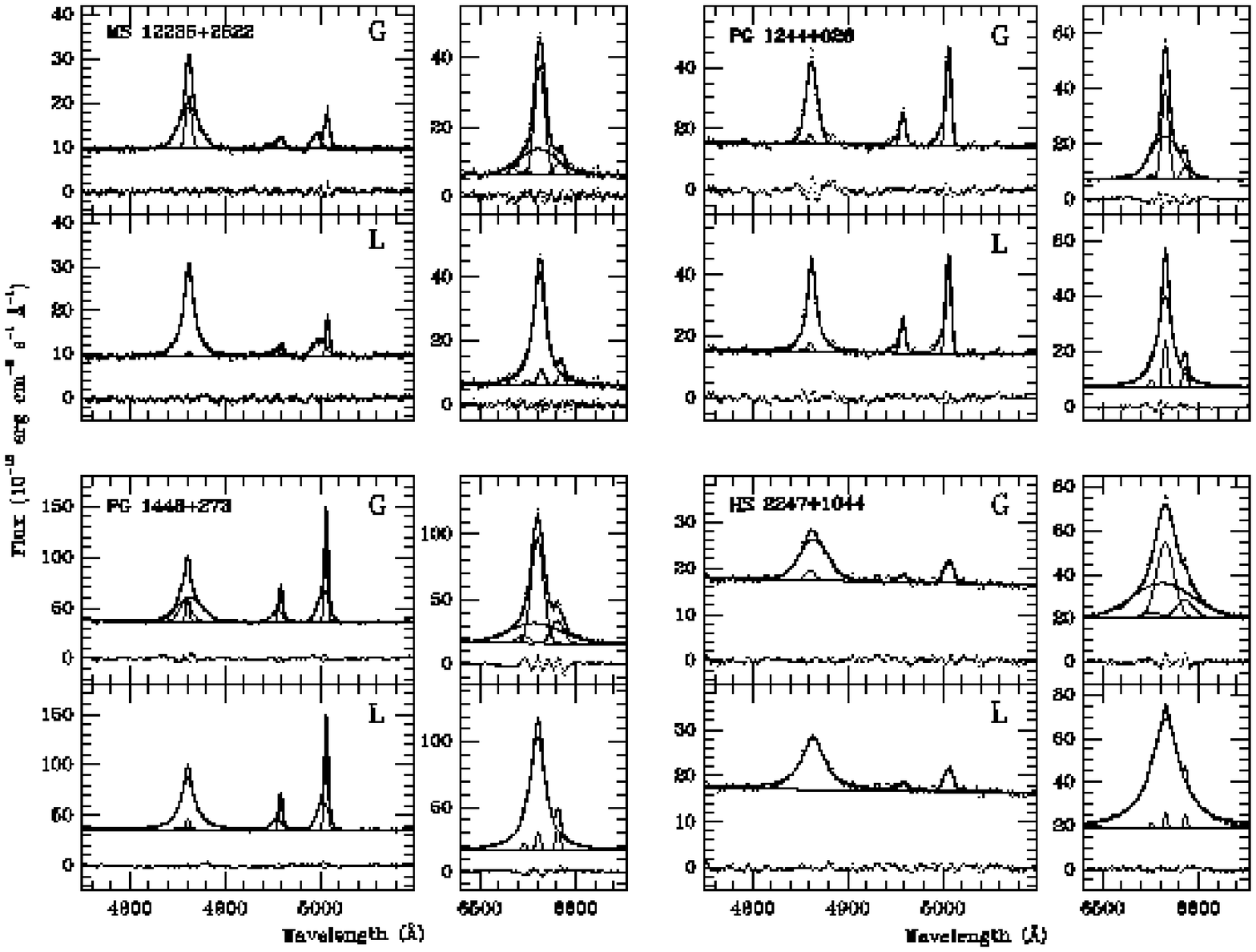}}
\caption{\label{Lorentzian_fit}
Deredshifted blue and red spectra of four NLS1 galaxies in our sample. The data 
points were 
fitted by a set of narrow Gaussian profiles, reproducing the narrow emission 
lines, plus a Gaussian (G) or Lorentzian (L) profile, to fit the broad Balmer 
component. The individual profiles are represented by a thin solid line, the 
total fit by a thick solid line and the differences between the data and the 
fit (the residuals) by the lower dotted line. 
}
\end{figure*}

 Although not all broad Balmer components in our sample are well fitted by a 
single Lorentzian, this is true in most cases (Table \ref{fwhm} lists, for all 
observed objects, the FWHM of the H$\alpha$ and H$\beta$ broad components 
obtained by using a single Lorentzian). As an illustration, we show in Fig.
\ref{Lorentzian_fit} the blue and red spectra of four objects fitted by a single
broad Gaussian and by a single broad Lorentzian; the narrow lines have been 
fitted either by a single Gaussian or by two Gaussians.
 In three cases, the fit of the red spectra is significantly better with a 
Lorentzian than with a Gaussian, as judged from the residuals. The improvement 
of the fit of the blue spectra with a Lorentzian is significant only for 
PG~1244$+$026. But, in all cases, the Lorentzian fit leads to 
$\lambda$6583/H$\alpha$ and $\lambda$5007/H$\beta$ ratios which are more similar
to the values expected for Seyfert 2 nebulosities (see below).

 The use of a Lorentzian rather than a Gaussian leads, for the 
NLR of most NLS1s, to line ratios typical of classical Seyfert 2s which strongly 
confirms that these broad Balmer components have a profile which is genuinely 
more similar to a Lorentzian.
 We give in Table \ref{FWRAP} the FWHM of the broad Gaussian and Lorentzian 
profiles for these four objects; we first note that the FWHM of the H$\alpha$ 
and H$\beta$ Lorentzians are equal within $\sim$10\%, while the H$\alpha$ FWHM 
of the Gaussians are on average 50\% larger than the H$\beta$ FWHM; this 
probably reflects the better quality of the fits obtained with Lorentzians. The 
FWHM of the Gaussians are systematically larger than the FWHM of the 
Lorentzians; in particular, HS~2247$+$1044 would not be classified as an NLS1 
from a Gaussian fit. \\

 The fact that the broad Balmer lines in NLS1s have a nearly Lorentzian profile 
seems to suggest that they are produced in a large disk without the contribution 
from an outflowing component. However, the high ionization UV emission lines in 
the spectrum of NLS1s appear to be significantly broader than H$\beta$ 
(Rodr\'{\i}guez-Pascual et al. 1997; Kuraszkiewicz et al. 2000; Wills et al. 
2000). In addition, Leighly (2001) showed that the high-ionization UV emission 
lines in two NLS1s (IRAS~13224$-$3809 and 1H~0707$-$495) are much broader and 
strongly blueshifted than the low-ionization lines. These observations suggest 
that, in NLS1s, the outflowing component exists but is relatively weak, its 
contribution being negligible in the optical range.
 

\begin{table*}
\caption{\label{fwhm} Line properties of the 59 
observed NLS1s (H II components have been systematically removed). Col. 1: 
name, col. 2: R$_{4570}$, ratio of the Fe II to the total H$\beta$ fluxes, 
cols. 3 and 4: FWHM (in km s$^{-1}$) of the broad 
H$\alpha$ and H$\beta$ lines fitted with a Lorentzian profile, col. 5:
EW (\AA) of the broad H$\beta$ component, col. 6: 
R$_{5007}$, ratio of the $\lambda$5007 to the total H$\beta$ fluxes, col 7:
FWHM (corrected for instrumental broadening) of the [O III] line, cols. 8 and 
9: FWHM of the individual components (F1 and F2) of the [O III] lines, col. 
10: velocity difference between the two [O III] components (broad--narrow), 
col. 11: intensity ratio of the two [O III] components (broad/narrow). 
a:   a good fit can only be obtained with two broad components, a Lorentzian 
and a Gaussian or two Lorentzians. 
b:   the broad H$\beta$ component is weak. } 
\begin{center}
\begin{tabular}{lrrrrrrrrrr}
\hline
Name & R$_{4570}$ & \multicolumn{2}{c}{FWHM (L)} & EW & R$_{5007}$ & FWHM & FW1 & FW2 & 
$\Delta$V & F2/F1  \\
   &  & H$\alpha$ & H$\beta$ &  H$\beta_{br}$ & & $\lambda$5007 & & & &   \\
\hline

 Mark\,335          &     0.38  &   a~~&   a~~&  86 & 0.25 & 280 &  245 &  910 &  $-$190 &  1.23 \\
 I\,Zw\,1           &     1.14  &   a~~&   a~~&  48 & 0.44 &1660 & 1040 & 1070 & $-$1180 &  0.43 \\
 Ton\,S180          &     1.03  & 1250 & 1085 &  46 & 0.18 & 675 &  435 & 1060 &  $-$465 &  1.00 \\
 Mark\,359          &  $<$0.5~~~&  830 &  900 &  18 & 1.32 & 180 &      &      &         &       \\
 MS\,01442$-$0055   &  $<$0.6~~~& 1260 & 1100 &  35 & 0.35 & 240 &      &      &         &       \\
 Mark\,1044         &     0.61  &  905 & 1010 &  63 & 0.15 & 420 &  335 &  720 &  $-$340 &  0.69 \\
 HS\,0328+0528      &  $<$0.6~~~& 1455 & 1590 &  79 & 1.67 & 220 &      &      &         &       \\
 IRAS\,04312$+$4008 &  $<$1.4~~~& 1060 &  860 &  16 & 0.35 & 380 &      &      &         &       \\
 IRAS\,04416$+$1215 &     1.14  & 1500 & 1470 &  55 & 0.67 &1320 &  650 & 1790 &  $-$480 &  1.95 \\
 IRAS\,04576$+$0912 &  $<$1.0~~~& 1100 & 1210 &  13 & 1.40 &1290 &  380 & 1260 &  $-500$ &  6.40 \\
 IRAS\,05262$+$4432 &     1.09  &  695 &  740 &  23 & 0.35 & 365 &      &      &         &       \\
 RX\,J07527$+$2617  &     0.71  & 1080 & 1185 &  50 & 0.21 & 400 &      &      &         &       \\
 Mark\,382          &  $<$0.8~~~& 1270 & 1280 &  20 & 0.56 & 155 &      &      &         &       \\
 Mark\,705          &     0.36  & 1745 & 1790 &  85 & 0.41 & 365 &  365 & 1630 &  $-$290 &  0.62 \\
 Mark\,707          &     0.47  & 1180 & 1295 & 102 & 0.53 & 315 &      &      &         &       \\
 Mark\,124          &     0.60  & 1645 & 1840 &  43 & 0.72 & 540 &  380 &  935 &  $-$335 &  0.75 \\
 Mark\,1239         &     0.63  &  905 & 1075 &  78 & 1.29 & 630 &  400 & 1395 &  $-$475 &  1.16 \\
 IRAS\,09571$+$8435 &     1.05  & 1270 & 1185 &  26 & 0.53 & 430 &  240 &  845 &  $-$370 &  1.07 \\
 PG\,1011$-$040     &     0.46  & 1370 & 1455 &  41 & 0.33 & 400 &      &      &         &       \\
 PG\,1016$+$336     &     0.60  & 1205 & 1590 &  65 & 0.06 & 315 &      &      &         &       \\
 Mark\,142          &     0.92  & 1335 & 1370 &  60 & 0.14 & 260 &      &      &         &       \\
 KUG\,1031$+$398    &  $<$1.5~~~& 1225 &  935 &  17 & 0.93 &1000 &  315 & 1115 &  $-$280 &  1.27 \\
 RX\,J10407$+$3300  &  $<$0.6~~~& 1425 & 1985 &  32 & 0.75 & 460 &      &      &         &       \\
 Mark\,734          &     0.67  & 1345 & 1825 &  45 & 0.49 & 450 &  180 &  525 &  $-$230 &  1.41 \\
 Mark\,739E         &     0.61  & 1415 & 1615 &  63 & 0.23 & 380 &      &      &         &       \\
 MCG\,06.26.012     &     0.52  & 1070 & 1145 &  83 & 0.14 & 220 &      &      &         &       \\
 Mark\,42           &     0.90  &  805 &  865 &  36 &      & 220 &      &      &         &       \\
 NGC\,4051          &     0.25  &  970 & 1120 &  50 & 0.55 & 300 &  200 &  665 &  $-$175 &  0.65 \\

\hline
\end{tabular}
\end{center}
\end{table*}
\addtocounter{table}{-1}
\begin{table*}
\caption{(end)}
\begin{center}
\begin{tabular}{lrrrrrrrrrr}
\hline
Name & R$_{4570}$ & \multicolumn{2}{c}{FWHM (L)} & EW & R$_{5007}$ & FWHM & FW1 & FW2 & 
$\Delta$V & F2/F1  \\
   &  & H$\alpha$ & H$\beta$ &  H$\beta_{br}$ & & $\lambda$5007 & & & &   \\
\hline

 PG\,1211$+$143     &     0.44  & 1400 & 1975 &  95 & 0.12 & 510 &      &      &         &       \\
 Mark\,766          &     0.35  & 1150 & 1630 &  55 & 1.83 & 330 &  220 &  710 &   $-$90 &  0.73 \\
 MS\,12170$+$0700   &     0.65  & 1405 & 1765 &  79 &      & 365 &      &      &         &       \\
 MS\,12235$+$2522   &     0.62  &  705 &  800 &  43 & 0.24 & 875 &  240 &  905 &  $-$570 &  1.34 \\
 IC\,3599           &           &  500 & --~~ &   6 & 3.23 & 280 &      &      &         &       \\
 PG\,1244$+$026     &     1.09  &  820 &  740 &  36 & 0.47 & 415 &  330 &  740 &  $-$390 &  0.40 \\
 NGC\,4748          &     0.55  & 1400 & 1565 &  65 & 1.34 & 365 &  295 & 1170 &  $-$150 &  0.49 \\
 Mark\,783          &  $<$0.5~~~& 1510 & 1655 &  42 & 2.29 & 430 &      &      &         &       \\
 R\,14.01           &     0.44  & 1470 & 1605 &  85 & 0.28 & 430 &      &      &         &       \\
 Mark\,69           &     0.59  & 1445 & 1925 &  34 &      & 315 &      &      &         &       \\
 2E\,1346$+$2646    &           & 1235 &   b~~&     & 2.40 & 330 &  180 &  950 &  $-$105 &  1.07 \\
 PG\,1404$+$226     &     0.85  & 1015 & 1120 &  65 & 0.19 & 950 &      &      &         &       \\
 Mark\,684          &     0.91  &   a~~& 1150 &  40 & 0.14 &1290 &      &      &         &       \\
 Mark\,478          &     0.55  & 1190 & 1270 &  72 & 0.17 & 920 &  365 & 1230 &  $-$475 &  2.05 \\
 PG\,1448$+$273     &     0.73  &  915 & 1050 &  36 & 0.61 & 315 &  155 &  890 &  $-$215 &  1.11 \\
 MS\,15198$-$0633   &           & 1115 &  --~~&  -- &      &     &      &      &         &       \\
 Mark\,486          &     0.46  & 1400 & 1680 & 123 & 0.13 & 400 &      &      &         &       \\
 IRAS\,15462$-$0450 &     0.59  & 1830 & 1615 &  41 & 0.62 &1600 &      &      &         &       \\
 Mark\,493          &     0.87  &  870 &  740 &  46 & 0.26 & 450 &  315 &  845 &  $-$400 &  0.75 \\
 EXO\,16524$+$3930  &  $<$0.9~~~& 1025 & 1355 &  39 & 0.21 & 400 &      &      &         &       \\
 B3\,1702$+$457     &  $<$1.2~~~&  930 &  975 &  19 & 2.03 & 365 &  295 & 1200 &  $-$280 &  0.56 \\
 RX\,J17450$+$4802  &     0.78  &  --~~& 1355 &  48 & 0.45 & 400 &      &      &         &       \\
 Kaz\,163           &     0.35  & 1325 & 1875 &  87 & 0.70 & 480 &      &      &         &       \\
 Mark\,507          &     1.94  & 1205 & 1565 &   6 & 0.50 &1025 &      &      &         &       \\
 HS\,1817$+$5342    &     0.59  & 1625 & 1615 &  79 & 0.20 &1000 &  570 & 1215 &  $-$375 &  2.21 \\
 HS\,1831$+$5338    &     0.74  & 1470 & 1555 &  32 & 0.30 & 240 &      &      &         &       \\
 Mark\,896          &     0.50  & 1015 & 1135 &  32 & 0.19 & 315 &      &      &         &       \\
 MS\,22102$+$1827   &  $<$1.2~~~&  820 &  690 &  42 & 0.16 & 890 &      &      &         &       \\
 Akn\,564           &     0.67  &  710 &  865 &  48 & 0.92 & 220 &      &      &         &       \\
 HS 2247$+$1044     &     1.11  & 1625 & 1790 &  30 & 0.12 & 710 &      &      &         &       \\
 Kaz 320            &     0.49  & 1160 & 1470 &  85 & 0.70 & 350 &  260 &  830 &  $-$275 &  0.57 \\

\hline
\end{tabular}
\end{center}
\end{table*}

\subsection{The narrow line region}

\subsubsection{The line ratios of the narrow line region}
 
\begin{figure}[h]
\resizebox{8.8cm}{!}{\includegraphics{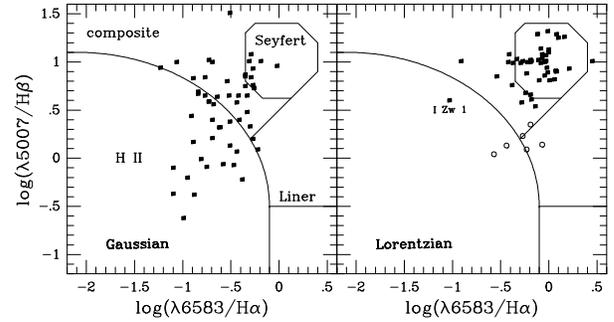}}
\caption{\label{diag} 
Diagnostic diagrams for the narrow line region of the objects in our sample. In
the left panel, the line ratios have been derived by fitting the broad Balmer 
components with a single Gaussian, while in the right panel we have used a 
single Lorentzian. In the right panel, Seyfert 2s are shown by black squares and
H II regions by open circles.} 
\end{figure}

\begin{figure}[h]
\resizebox{8.8cm}{!}{\includegraphics{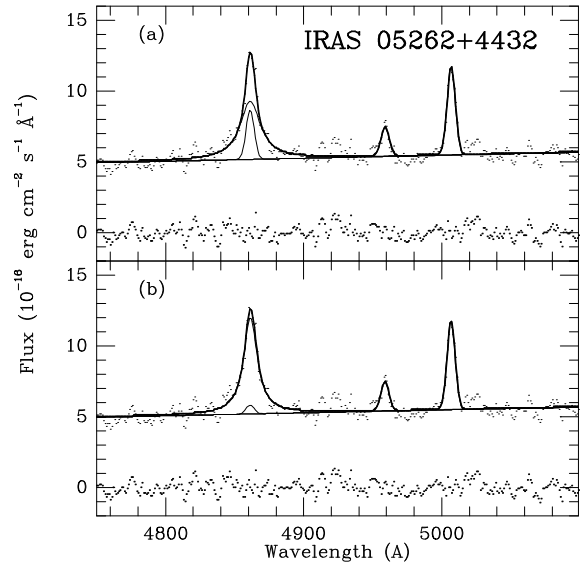}}
\caption{\label{i05262} In (a), we have fitted the blue spectrum of 
IRAS 05262$+$4432
with a set of Gaussian profiles for the narrow lines plus a single Lorentzian
profile for the broad component of H$\beta$; in (b), we forced the
$\lambda$5007/H$\beta$ ratio to be equal to 10. Examination of the residuals 
shows that the two solutions have equal merits.}
\end{figure}

 As we have seen above, for all objects for which a single Lorentzian gave a 
good fit to the broad Balmer component, we also made a fit with a single 
Gaussian. In this way, we obtained two sets of line ratios 
($\lambda$5007/H$\beta$ and $\lambda$6583/H$\alpha$) which are plotted in Fig. 
\ref{diag} (when the signal to noise ratio is low, the fit with a single broad 
H$\beta$ Lorentzian sometimes gives a low value for the $\lambda$5007/H$\beta$ 
ratio; in these cases, when forcing this ratio to be equal to 10, the fit is 
not worse as illustrated for IRAS~05262+4432 in Fig. \ref{i05262}). The left 
panel shows the ratios obtained by fitting the broad Balmer component with a
Gaussian; the right panel, the ratios obtained with a Lorentzian. While, in the 
left panel, many points are located in the H II region or composite domain
(Gon\c{c}alves et al. 1999a), in the right panel, most of the points are in the 
Seyfert 2 region; one object (HS~0328$+$0528) belongs to the rare class of 
weak-[N II] Seyfert galaxies (Gon\c{c}alves et al. 1999a) while six (Mark~42, 
MS~12170$+$0700, Mark~69, Mark~684, IRAS~15462$-$0450 and Mark~507) have an 
emission line component which falls in or near the H II region domain and is a 
genuine H II region; they are shown as open circles; these objects are most 
probably composite type objects similar to those described by V\'eron et al. 
(1981b) and Gon\c{c}alves et al. (1999a). 
 The fit of the broad component of I~Zw~1 with a single component is clearly
unsatisfactory; we got a better fit with two Lorentzians which puts the 
representative point of the narrow line region in the Seyfert 2 area.

\begin{table}[!]
\caption{\label{FWRAP}FWHM of the broad component of the H$\alpha$ and H$\beta$
lines when fitted either by a Gaussian or a Lorentzian profile. Col. 
1: name, cols. 2 and 3: FWHM (km s$^{-1}$) of the broad H$\alpha$ and H$\beta$ 
lines fitted 
with a Gaussian, col. 5: ratio (R) of the FWHM of H$\alpha$ to H$\beta$, cols. 
6 and 7: FWHM (km s$^{-1}$) of the broad H$\alpha$ and H$\beta$ lines fitted 
with a Lorentzian, col. 8: ratio of the FWHM of H$\alpha$ to H$\beta$. }
\begin{center}
\begin{tabular}{l|rrc|rrc}
\hline
  &  \multicolumn{3}{c|}{Gaussian fit}    & \multicolumn{3}{c}{Lorentzian fit}\\
 Name & H$\alpha$ & H$\beta$ & R & H$\alpha$ & H$\beta$ & R  \\
  &  \multicolumn{2}{c}{FWHM} & & \multicolumn{2}{c}{FWHM} & \\
\hline
MS 12235$+$2522   & 2100 & 1585 & 1.32 &  705 &  800 & 0.88 \\
PG 1244$+$026     & 1625 &  945 & 1.72 &  820 &  740 & 1.11 \\
PG 1448$+$273     & 3000 & 1890 & 1.59 &  915 & 1050 & 0.87 \\
HS 2247$+$1044    & 3410 & 2280 & 1.50 & 1625 & 1790 & 0.91 \\
 
\hline
\end{tabular}
\end{center}
\end{table}
\normalsize

 In contrast, Rodr\'{\i}guez-Ardila et al. (2000b) found that the 
$\lambda$5007/H$\beta_{\rm n}$ ratio emitted in the NLR of NLS1s varies from 1 
to 5, instead of $\sim$10 for BLS1s. This is most probably due to the fact 
that they modelled the broad Balmer component with a Gaussian rather than a 
Lorentzian. On the other hand, Nagao et al. (2001) showed that the line ratios 
[O I]$\lambda$6300/[O III]$\lambda$5007 and 
[O III]$\lambda$4363/[O III]$\lambda$5007 are statistically indistinguishable in
NLS1s and BLS1s.

\subsubsection{The kinematics of the narrow line region}

 The majority of the profiles of the [O III] lines in Seyfert 1s are markedly 
asymmetric, exhibiting a sharper falloff to the red than to the blue (Heckman et
al. 1981; Vrtilek \& Carleton 1985; Whittle 1985a; Veilleux 1991b); these 
profiles often have a two component structure with a narrow core superposed on 
a broader, blueshifted base (Heckman et al. 1981; Leighly 1999b); it seems that 
the velocity of the core is equal to the systemic velocity of the galaxy 
(Heckman et al. 1981; 1984; Whittle 1985a). 

 Many Seyferts have an ``ionization cone" appearing as either one- or two-sided 
structures emanating from the active nucleus; when single (one-sided) cones are 
seen, they generally project against the far side of the galaxy disk, suggesting
that the counter cone is present but obscured by dust in the disk (Pogge 1989; 
Wilson et al. 1993; Wilson 1994).
 Colbert et al. (1996) suggested that more than 25\% of Seyfert galaxies have 
good evidence (from the kinematics of the H$\alpha$ and [N II] lines) for minor 
axis galactic outflows.
 Crenshaw et al. (1999) showed, from the observation of UV absorption lines
(mainly C IV and N V), that large scale outflows with a velocity of a few 
hundreds km s$^{-1}$ are common in Seyfert galaxies.  High spatial resolution
spectroscopic observations of the nuclear emission lines of NGC~1068 (Crenshaw
\& Kraemer 2000) and NGC~4151 (Crenshaw et al. 2000; Kaiser et al. 2000) give 
strong support for biconical radial outflows in these objects.

 Whittle et al. (1988), Storchi-Bergmann et al. (1992), Arribas \& Mediavilla
(1993) and Kaiser et al. (2000) showed that the AGN NLRs have two line-emitting 
constituents: the 
``[O III] components" (the broad blueshifted bases of the [O III] lines; Heckman
et al. 1981; Vrtilek 1985; Whittle 1992; Christopoulou et al. 1997), to be 
identified with the biconical structure and which are outflowing from the 
nucleus with a speed of a few hundreds km s$^{-1}$, and the ``ambient [O III] 
emission" which is more widely distributed and shows normal galactic rotation.
 A detailed kinematic study of the emission line gas in the Seyfert galaxy 
NGC 2992 shows the presence of both a disk component which is well modeled
by pure circular rotation and an outflowing component distributed into two wide 
cones (Veilleux et al. 2001). 
 Smith (1993) suggested that the ``[O III] component" could be accelerated 
outwards by a supersonic wind generated by the active nucleus. The relative 
intensity of the two components varies widely from object to object. 

 High-ionization lines ([Fe VII], [Fe X] and [Fe XI]) are often present in 
emission in NLS1s; these coronal lines tend to be blueshifted relative to, and
broader than, the low-ionization lines; the systematic blueshift indicates an 
outflow of the gas emitting these features (Grandi 1978; Erkens et al. 1997). 

 There is a strong correlation between the [O III]$\lambda$5007 line width and 
luminosity (Whittle 1985b).
 The [O III] FWHM is correlated with the absolute magnitude of the galaxy bulge 
(V\'eron \& V\'eron-Cetty 1986) and the observed galaxy rotation (V\'eron 
1981; Whittle 1992) showing that gravity plays a dominant role in the NLR of 
most objects, at least in the ``ambient [O III] emission" region which dominates
in most AGNs. \\
 
  For a number of objects in our sample (30), the fit of the [O III] lines with 
a single Gaussian was poor, either because of the simultaneous presence on the 
slit of a H II region (6), or because the lines are asymmetrical with a blue 
wing (24); in these last cases, we fitted them with two Gaussians  following the
procedure described in section~2; as a result, we found a red component with a 
FWHM in the range 180--650 km s$^{-1}$ and a blue component with a FWHM in the 
range 525--1\,790 km s$^{-1}$, blueshifted by 90 to 570 km s$^{-1}$ with respect
to the red component (Table \ref{fwhm};  in this table, the [O III] FWHM have 
been corrected for the instrumental profile); there is one exception, I~Zw~1, 
for which the red component is exceptionally broad (1\,040 km s$^{-1}$ FWHM) and
the blue component is blueshifted by 1\,180 km s$^{-1}$. This is in agreement 
with previous findings. 
  
  For the objects requiring an additional [O III] component, the associated
H$\beta$ line was often very weak; as a result, the fitting routine sometimes 
gave a negative flux; in such cases, we set the ratio $\lambda$5007/H$\beta$
to 10, the mean value for Seyfert galaxies. This is a common procedure used to
prevent the fitting routine to yield non-physical values whenever the line
intensities are too small to be disentangled from the noise. \\

 The soft X-ray photon index is available for 41 of the objects for which we 
have a
blue spectrum. Twenty of them have $\Gamma>2.9$, 21 have a smaller value; 
among the 20 objects with a steep soft X-ray spectrum, 12 have a blueshifted 
[O III] component, while only eight of the other set have such a component; 
moreover, two of the eight steep X-ray spectrum objects without a blueshifted
component (PG~1404+226 and RXS~J07527+2617) have an optical spectrum with a 
poor signal to noise ratio and relatively broad [O III] lines (950 and 400 km 
s$^{-1}$ respectively) so that the presence of a blueshifted component could 
have been overlooked.  So there is a weak, statistically unsignificant, trend
for the objects with a soft X-ray excess to have an outflowing [O III] 
component. It is interesting to remark that Erkens et al. (1997) have found that 
strong coronal lines occur predominantly in objects with the steepest soft X-ray
spectra and that these lines are relatively broad and bluehifted. The FWHM and
velocity shifts are comparable for the blueshifted [O III] lines and the coronal
lines. Could these lines all come from the same emitting clouds?

\subsection{The Fe II emission}

\subsubsection{Correlation between the strength of the Fe II emission and the 
width of the broad H$\beta$ component}

 Typical AGNs have R$_{4570}\sim$0.4 with $\sim$90\% of objects in the range 0.1
to 1 (Osterbrock 1977a; Bergeron \& Kunth 1984). Moderately strong Fe II 
emission (R$_{4570}$ $>$ 1) occurs in perhaps 5\% of all objects (Lawrence et 
al. 1988). A few superstrong Fe II emitters (R$_{4570}$ $>$ 2) have been found 
(Lawrence et al. 1988; Lipari et al. 1993; Lipari 1994; Moran et al. 1996; Xia 
et al. 1999); they are listed in Table \ref{superfe}. They are roughly an order
of magnitude rarer.

\begin{table}[h]
\caption{\label {superfe}Known superstrong Fe II emitters (R$_{4570} >$ 2)} 
\begin{center}
\begin{tabular}{llcc}
\hline
Name & Position & R$_{4570}$ & H$\beta_{\rm br}$ FWHM \\
     &          &       & km s$^{-1}$ \\
\hline
 IRAS 04312$+$4008  & 0431$+$40  & 2.4 & 1230 \\
 IRAS 07598$+$6508  & 0759$+$65  & 2.6 & 3200 \\
 IRAS 10026$+$4347  & 1002$+$43  & 2.0 & 2500 \\
 IRAS 11598$-$0112  & 1159$-$01  & 3.3 & \verb+ +780 \\
 Mark 231           & 1254$+$57  & 2.1 & 3000 \\
 IRAS 13224$-$3809  & 1322$-$38  & 2.4 & \verb+ +650 \\
 Mark 507           & 1748$+$68  & 2.9 & \verb+ +965 \\
 IRAS 18508$-$7815  & 1850$-$78  & 2.4 & 3100 \\
 IRAS 23410$+$0228  & 2341$+$02  & 4.0 & \verb+ +970 \\
\hline
\end{tabular}
\end{center}
\end{table}
\normalsize

 Wills (1982) was the first to suggest that R$_{4570}$ is roughly inversely 
proportional to line width (FWHM of the broad H$\beta$ component). Gaskell 
(1985) showed that R$_{4570}$ increases dramatically for FWHM $<$ 1\,600
km s$^{-1}$, but is relatively constant for FWHM $>$ 1\,600 km s$^{-1}$. Zheng 
\& Keel (1991) found that for AGNs with FWHM $>$ 6\,000 km s$^{-1}$, the mean 
value of R$_{4570}$ is 0.21, less than half of that of the other objects, 
confirming that strong Fe II emission is not found in objects showing very broad
emission lines; they showed that this is not an artifact resulting from blending
of the Fe II lines when they are broad. Boroson \& Green (1992), followed by 
Wang et al. (1996) and Rodr\'{\i}guez-Ardila et al. (2000a), confirmed the 
existence of a strong anticorrelation between R$_{4570}$ and H$\beta$ FWHM. This
anticorrelation could be due either to the existence of an anticorrelation
between Fe II EW and H$\beta$ FWHM (Zheng \& O'Brien 1990; Boroson \& Green 
1992) or of a correlation between H$\beta$ EW and FWHM (Osterbrock 1977a; 
Gaskell 1985; Goodrich 1989).

\begin{figure}[h]
\resizebox{8.8cm}{!}{\includegraphics{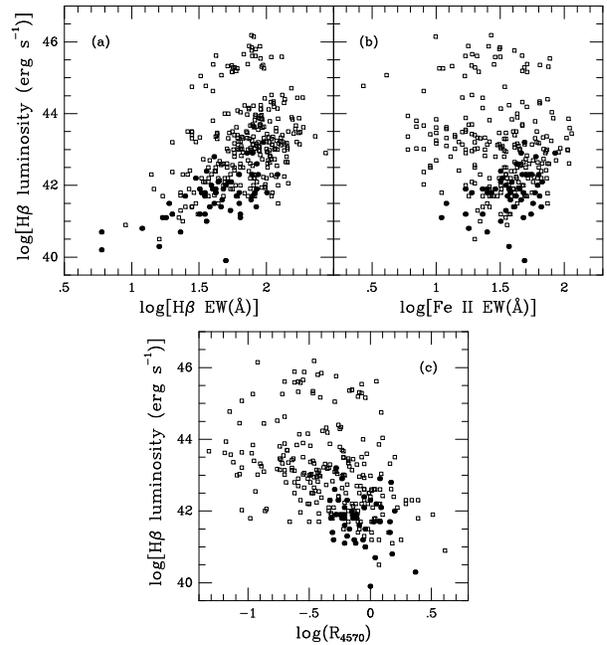}}
\caption{\label{r4570} Plots of the broad H$\beta$ luminosity: in (a) {\it vs} 
the EW of the broad H$\beta$ component; the two parameters are highly correlated
 (r=0.43) for low luminosity objects (L(H$\beta)<$10$^{42.5}$ erg s$^{-1}$); but 
the H$\beta$ EWs saturate at $\sim$100 \AA\ for larger luminosity objects; in 
(b), {\it vs} the Fe II EW; it is a scatter diagram  (r=0.20); in (c) 
{\it vs} R$_{4570}$. The anticorrelation observed (r=0.42) is 
clearly due to the weakness of H$\beta$ in the low luminosity objects rather 
than to the strength of 
R$_{4570}$. Several samples of Seyfert 1s have been used; our own is shown as 
filled circles; others (see text) are shown as open squares.}
\end{figure}

 We have plotted in Fig. \ref{r4570} the broad H$\beta$ luminosity {\it vs} the
H$\beta_{\rm br}$ EW, the Fe II EW and R$_{4570}$ for all objects in our sample 
and in those of Boroson \& Green (1992), Brotherton (1996), Moran et al. (1996),
Marziani et al. (1996), Corbin \& Boroson (1996), Corbin (1997), Grupe et al. 
(1999) and McIntosh et al. (1999); Grupe et al. reported the EW of the 
whole Fe II complex from 4250 to 5880 \AA; Leighly (1999b) used 30\% of this 
value; we used 20\%. This figure clearly shows that the anticorrelation between
H$\beta$ luminosity and R$_{4570}$  (r=0.42) is the result of the correlation 
between H$\beta$ luminosity and EW  (r=0.43) as there is no correlation between 
H$\beta$ luminosity and Fe II EW  (r=0.20) (here and in what follows, r is the 
linear correlation coefficient; the number of degrees of freedom is in each case
near 290). This confirms Gaskell (1985) result that NLS1s have weak H$\beta$ 
rather than strong Fe II.
 At very high densities (N$_{e}>$10$^{10}$ cm$^{-3}$), the hydrogen lines become
thermalized and their intensity drops considerably (Rees et al. 1989) which is a
possible explanation of the decrease of the H$\beta$ EW in NLS1s (Gaskell 1985).

\subsubsection{Correlation between the strength of the Fe II and [O III] 
emissions}

 Goodrich (1989) noted that a defining characteristic of NLS1s is that the ratio
R$_{5007}$ of the [O III]$\lambda$5007 flux to the total H$\beta$ flux is $<$3. 
In fact, when the first NLS1s were observed spectroscopically, 
the spectra were of relatively low resolution and the permitted and forbidden 
lines were believed to have the same width; these objects could therefore be 
mistaken for Seyfert 2s except for the presence in their spectra of strong Fe II
emission, a strong blue continuum and a small $\lambda$5007/H$\beta$ ratio due 
to the unrecognized presence of a broad H$\beta$ component; as, in Seyfert 2s, 
this ratio is larger than 3, a smaller value was an indication of the presence 
of a broad H$\beta$ component.
 But Osterbrock (1981) has divided the Seyfert 1s into five subgroups: S1.0, 
1.2, 1.5, 1.8 and 1.9 on the basis of the appearance of the Balmer lines; a
quantitative definition of these subgroups has been given by Winkler (1992) 
using the value of R$_{5007}$: for S1.0s,
R$_{5007}<$0.2, for S1.2, 0.2$<$R$_{5007}<$0.5, for S1.5, 0.5$<$R$_{5007}<$3.
The higher values of R$_{5007}$ observed in S1.8 and S1.9 are believed to be
due to partial extinction of the broad H$\beta$ component. It follows that, for
all S1s which do not suffer extinction, the condition R$_{5007}<$3 is fulfilled.
We have measured on our spectra the parameter R$_{5007}$ (given in Table 
\ref{fwhm}); we have excluded for the computation of R$_{5007}$ the contribution
of the H II regions mentioned above. All objects have R$_{5007}$ values lower 
than 0.8 except IC~3599 for which it exceeds~3.

 Boroson \& Green (1992), Grupe et al. (1999) and McIntosh et al. (1999) found 
that Fe II is strong in objects with weak [O III] lines and vice versa. Let us 
note however that while Grupe et al. and McIntosh et al. found an 
anticorrelation between R$_{5007}$ and Fe II EW, Boroson \& Green found a strong
anticorrelation, not between the Fe II and [O III] EWs, but rather between the 
Fe II EW and the ratio of the peak height of the [O III]$\lambda$5007 line to
that of H$\beta$ which depends both on the value of R$_{5007}$ and the width
of H$\beta$; this correlation results from the anticorrelation between 
R$_{4570}$ and the H$\beta$ FWHM. 

\begin{figure}[h]
\resizebox{8.8cm}{!}{\includegraphics{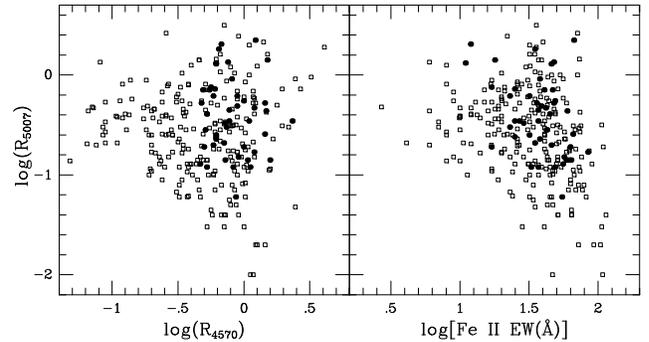}}
\caption{\label{o5007} Plot of R$_{5007}$ {\it vs} R$_{4570}$ (left panel) and 
Fe II EW (right panel) for several samples of Seyfert 1s. There is no 
correlation between R$_{5007}$ and R$_{4570}$  (r=0.05), while there is a trend 
for objects with a large Fe II EW to have a weak R$_{5007}$  (r=0.29). The 
symbols are as in Fig. 10.
}
\end{figure}

 In Fig. \ref{o5007}, we have plotted  R$_{5007}$ {\it vs} R$_{4570}$ and Fe II 
EW. There is a weak anticorrelation between Fe II EW and R$_{5007}$, but not 
between R$_{4570}$ and R$_{5007}$. Therefore we do not confirm that Fe II is 
strong when the [O III] lines are weak. 

\begin{figure}[h]
\resizebox{8.8cm}{!}{\includegraphics{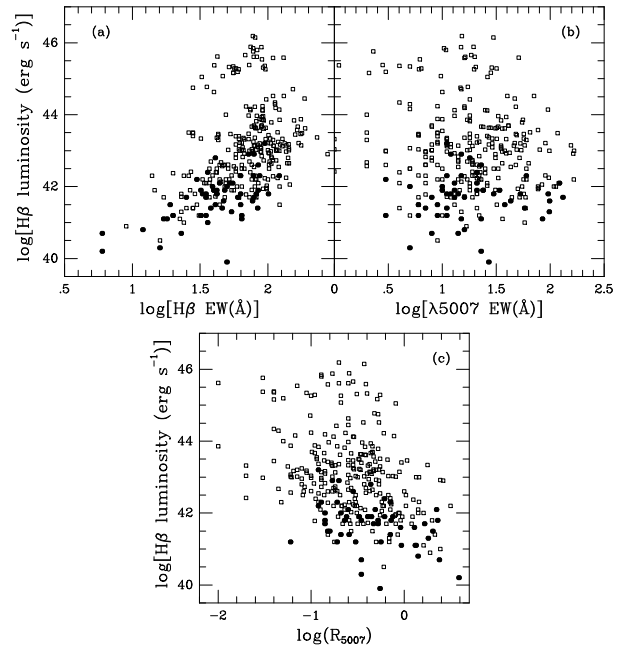}}
\caption{\label{hbetalum} Plots of the broad H$\beta$ luminosity: in (a) {\it 
vs} the EW of the broad H$\beta$ component (as in fig. 10); in (b), {\it vs} the
$\lambda$5007 EW. It is a scatter diagram  (r=0.16). As 
a result, the H$\beta$ luminosity is anticorrelated with R$_{5007}$ 
 (r=0.43) (c). 
The symbols are as in Fig. 10.}
\end{figure}

 Figure \ref{hbetalum} shows the broad H$\beta$ luminosity {\it vs} the broad 
H$\beta$ and $\lambda$5007 EWs, and R$_{5007}$; the H$\beta$ luminosities and 
$\lambda$5007 EWs are not correlated; the weak anticorrelation between H$\beta$
luminosity and R$_{5007}$ is due to the weakness of the H$\beta$ EW in low 
luminosity objects, rather than to the strength of the $\lambda$5007 EW.

\subsection{On the definition of NLS1s}

 Initially, as we have seen, NLS1s were defined as having narrow ``broad" Balmer
components ($<$2\,000 km s$^{-1}$ FWHM). However, there is a continuous 
distribution of optical line widths in Seyfert 1s and the separation between 
BLS1s and NLS1s is arbitrary (Turner et al. 1999b; Sulentic et al. 2000a). 
Sulentic et al. (2000a) suggested that Seyfert 1s with H$\beta$ FWHM $<$4\,000 
km s$^{-1}$ constitute a homogeneous class of objects having strikingly 
different line profiles compared to Seyfert 1s with broader lines. Grupe
et al. (1999) have found objects displaying NLS1 properties (strong Fe II 
emission, a soft X-ray excess and variability) in spite of their H$\beta$ FWHM 
exceeding 2\,000 km s$^{-1}$. IRAS~10026$+$4347 has a large soft X-ray excess 
($\Gamma$=3.2$\pm$0.5 or 2.9$\pm$0.2), strong Fe II emission (R$_{4570}$=2.0), 
and a high amplitude X-ray variability ($\times$8); but the H$\beta$ FWHM is 
$\sim$ 2\,500 or 2\,990 km s$^{-1}$ (Grupe et al. 1998; Xia et al. 1999). 
Similarly, PDS~456 has a H$\beta$ FWHM equal to 3\,500 km s$^{-1}$ although 
the Fe II lines are relatively strong (R$_{4570}$=0.46) (Simpson et al. 1999) 
and the X-ray spectrum shows a soft excess ($\Gamma$=3.9$\pm$0.8) (Reeves et al.
2000; Vignali et al. 2000). A few objects are known which have strong Fe II 
emission and relatively broad H$\beta$ lines, but no soft X-ray excess; Mark~231
and IRAS~07598$+$6508 are two such examples.
 Mark~231 has a strong Fe II emission (R$_{4570}$=2.03 or 1.60), its H$\beta$ 
FWHM is $\sim$ 3\,000 km s$^{-1}$ (Boroson \& Meyers 1992; Lipari et al. 
1993); the hard X-ray spectrum is heavily attenuated making it difficult to 
detect the eventual presence of a soft X-ray component (Turner 1999). 
 IRAS~07598$+$6508 also has strong Fe II emission (R$_{4570}$=2.60) and a 
relatively broad H$\beta$ line (2\,550--3\,200 km s$^{-1}$ FWHM) (Lawrence et 
al. 1997; Lipari et al. 1993; Boroson \& Meyers 1992); it is probably a highly 
obscured X-ray source (Gallagher et al. 1999). \\

\begin{figure}[h]
\resizebox{8cm}{!}{\includegraphics{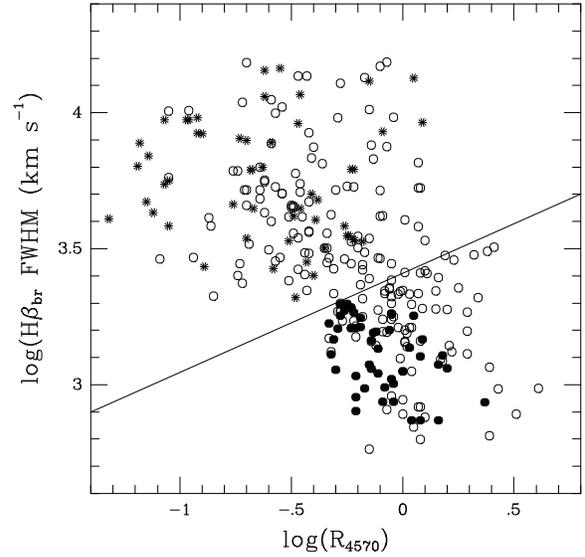}}
\caption{\label{FW_R4570} Plot of the broad H$\beta$ component FWHM
{\it vs} R$_{4570}$. Filled circles represent the objects in our 
sample, open circles other radio quiet objects and stars radio loud objects; the
correlation between the two parameters is strong  (r=0.54). The
straight line divides the plane into BLS1s (above the line) and what could be 
the newly defined class of NLS1s (below the line).}
\end{figure}

 We have plotted in Fig. \ref {FW_R4570}, H$\beta_{\rm br}$ FWHM {\it vs} 
R$_{4570}$ for all objects in our sample as well as in the other samples listed 
above; this plot shows a trend for the objects with the strongest Fe II emission
to have  narrower Balmer lines, with FWHM up to 3\,500 km s$^{-1}$. We have
drawn a line such that, for most objects below this line, R$_{4570}>$0.50 while,
for many objects above it, R$_{4570}<$ 0.50. This could be a better definition 
of NLS1s. \\

\begin{figure}[h]
\resizebox{8cm}{!}{\includegraphics{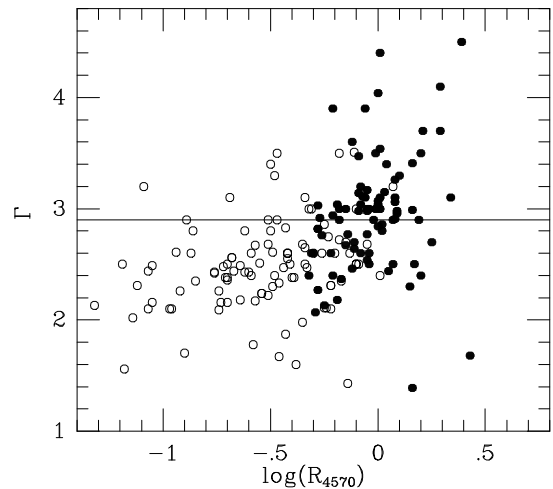}}
\caption{\label{gam_r4570} Plot of the {\it ROSAT} photon index $\Gamma$ {\it 
vs} R$_{4570}$. Filled circles represent the points below the line in the 
previous figure and open circles the points above the line. This diagram 
clearly shows that the soft X-ray excess tends to be large when the Fe II 
emission is large  (r=0.40).}
\end{figure}

 Figure \ref {gam_r4570} is a plot of the {\it ROSAT} photon index $\Gamma$ {\it
vs} R$_{4570}$; open 
circles represent the objects above the line in Fig. \ref {FW_R4570} and filled 
circles the objects below this line. This figure shows a definite correlation
between $\Gamma$ and R$_{4570}$. Although the objects with strong Fe II emission
(black dots) show a large dispersion in $\Gamma$, most objects with small Fe II
have $\Gamma<2.9$.

 Among the objects classified as NLS1s, having both narrow ``broad" Balmer 
components and a strong Fe II emission, two have a very small photon 
index: IRAS~09571+8435 ($\Gamma$=1.39) and Mark~507 ($\Gamma$=1.68). Mark~507 
has an intrinsic neutral hydrogen column density N$_{\rm H}$=27$\times$10$^{20}$
cm$^{-2}$ (see notes), this value being derived
assuming that the intrinsic X-ray spectrum is a single power-law; it is quite 
possible that the column density is even higher and hides a soft X-ray 
component; this could also be true for IRAS~09571+8435.

\begin{figure}[h]
\resizebox{8cm}{!}{\includegraphics{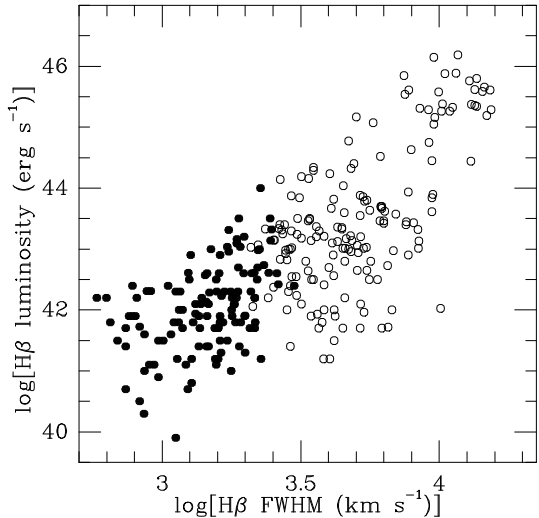}}
\caption{\label{hb_lfw} Plot of the broad H$\beta$ luminosity {\it vs} the 
FWHM of the broad H$\beta$ component. The symbols are as in Fig. 14. There is 
clearly a strong correlation between the two plotted parameters  (r=0.76).}
\end{figure}

 The FWHM of the broad H$\beta$ component has been found to increase with the 
H$\beta$ luminosity (Miller et al. 1992). Figure \ref {hb_lfw} is a plot of the
H$\beta$ luminosity {\it vs} the H$\beta$ FWHM; the correlation between these 
two parameters is indeed quite strong;  the linear correlation coefficient is 
r=0.76 for 294 data points, corresponding to a probability p$<$0.01\% for no 
correlation between the two parameters; luminous BLS1s tend to have
broader Balmer lines (but we should keep in mind that the published H$\beta$ 
FWHM have not been measured in a uniform way and that the presence of the narrow
component of the line has not always been taken properly into account). This 
leads to question whether the defining criterion for NLS1s should be a function 
of luminosity (Wills et al. 2000).

\section{Conclusions}

  In most cases, the broad Balmer components of NLS1s are well fitted by a 
single Lorentzian profile, confirming previous claims that Lorentzian rather 
than Gaussian profiles are better suited to reproduce the shape of the broad 
lines in NLS1s. This has consequences concerning their FWHMs and line ratios: 
when the broad Balmer components are fitted with a Lorentzian, most narrow line 
regions have line ratios typical of Seyfert 2s, while when they are fitted with
a Gaussian they are widely scattered in the usual diagnostic diagrams; moreover,
the FWHM of the best fitting Lorentzian is systematically smaller than the FWHM
of the Gaussian. The Lorentzian shape of the broad emission lines could be the 
result of these lines being produced in a large disk.

 We find that the [O III] lines usually have a relatively narrow Gaussian 
profile ($\sim$ 200--500 km s$^{-1}$ FWHM) with often, in addition, a second 
broad ($\sim$ 500--1\,800 km s$^{-1}$ FWHM), blueshifted Gaussian component. 

  We do not confirm that the [O III] lines are weak in NLS1s.

 As previously suggested, there is a continuous transition of all properties
between NLS1s and BLS1s. The strength of Fe II relative to H$\beta$ (R$_{4570}$)
could be the best parameter to recognize an NLS1 defined as a Seyfert 1 with a
high accretion rate. The maximum FWHM of the broad Balmer component of NLS1s 
could be an increasing function of the H$\beta$ luminosity.

 As all objects with H$\beta$ FWHM smaller than 2\,000 km s$^{-1}$ seem to be
genuine NLS1s, we plan to make similar spectroscopic observations of a sizeable
sample of Seyfert 1s with 2\,000 $<$ H$\beta$ FWHM $<$ 4\,000 km s$^{-1}$ in the
hope of better define the properties of those galaxies intermediate between
NLS1s and BLS1s. 

\section{Notes on individual objects}

\small
 {\bf Mark~335} (0003+19) is an NLS1 with strong Fe II emission and narrow 
``broad" Balmer lines (R$_{4570}$=0.62 and H$\beta$ FWHM=1\,640 km s$^{-1}$)
(Boroson \& Green 1992) and relatively strong coronal lines (Grandi 1978).

 The broad Balmer components clearly have a complex and asymmetrical profile 
(van Groningen 1987; Arav et al. 1997); to get a satisfactory fit of our 
spectra, two broad components were necessary, a Gaussian and a Lorentzian. \\

 {\bf I~Zw~1} (0050+12) is the prototype NLS1; its optical spectrum reveals 
narrow emission lines and strong Fe II emission (Phillips 1976). The H$\beta$
FWHM is in the range 1\,050--1\,500 km s$^{-1}$ (Osterbrock 1977a; Phillips 
1978a; Peterson et al. 1982; de Robertis 1985) although Wilkes et al. (1999) 
measured 2\,590 km s$^{-1}$.
 The broad Balmer components clearly have a complex and asymmetrical profile; to
get a satisfactory fit of our spectra, two Lorentzians were needed.

 An UV spectrum shows that ions of increasing ionization level show increasing 
excess blue-wing flux and an increasing line peak velocity shift (Laor et al. 
1997b).
 UV absorption lines have been observed, indicating the likely presence of a 
warm absorber (Crenshaw et al. 1999). \\

 {\bf Ton~S180} (0054$-$22) is an NLS1 (Comastri et al. 1998) with strong 
Fe II emission (R$_{4570}$=0.84, Comastri et al. 1998, or 1.02, Winkler 1992) 
and narrow Balmer lines (FWHM$\sim$1\,000 km s$^{-1}$) (Winkler 1992; Comastri 
et al. 1998; Grupe et al. 1999). \\
  
 {\bf Mark~359} (0124+18) is an NLS1 (Osterbrock \& Pogge 1985). The FWHM of 
the broad H$\alpha$ component is 780 km s$^{-1}$; the forbidden lines are 
exceptionally narrow (FWHM$\sim$100 km s$^{-1}$) (Veilleux 1991a). Discrepent 
values have been published for R$_{4570}$ : 0.62 (Goodrich 1989) and $<$0.08 
(Osterbrock \& Pogge 1985).
 The coronal lines are very strong (Veilleux 1988; Erkens et al. 1997). \\

 {\bf MS~01442$-$0055} (0144$-$00) is an NLS1 with H$\beta$ FWHM= 1\,940 km 
s$^{-1}$ and R$_{4570}$=0.78 (Stephens 1989). \\

 {\bf Mark~1044} (0227$-$09) is an NLS1 with strong narrow Fe II emission 
(Osterbrock \& Dahari 1983). The H$\beta$ FWHM is 1\,280 (Goodrich 1989) or 
1\,400 km s$^{-1}$ (Rafanelli 1985). \\

 {\bf HS~0328+0528} (0328+05). The spectra published by Perlman et al. (1996) 
and Engels et al. (1998) suggest that it is an NLS1, although they do not show 
the presence of Fe II emission. Our spectra show that the broad Balmer 
components are narrow ($\sim$ 1\,500 km s$^{-1}$) which confirms the 
classification as an NLS1.

 The [N II] lines are extremely weak; we measured $\lambda$6583/H$\alpha$=
0.12. \\

 {\bf IRAS~03450+0055} (0345+00). Boroson \& Meyers (1992) measured an 
H$\alpha$ FWHM of 1\,310 km s$^{-1}$ and R$_{4570}$=0.96. Giannuzzo \& Stirpe 
(1996) included it in their sample of NLS1s. 

 We have not observed this object. \\

 {\bf IRAS~04312+4008} (0431+40) is an NLS1 with very strong Fe II emission:
R$_{4570}$=2.36 (Moran et al. 1996).
 It is located at low galactic latitude: b=--5.0\degr\ with 
N$_{\rm H}\sim$3.4$\times$10$^{21}$ cm$^{-2}$. \\

 {\bf Mark~618} (0434$-$10). The Balmer line FWHM lies in the range 
1\,760--2\,300 km s$^{-1}$ (Osterbrock 1977a; Feldman \& MacAlpine 1978; 
Phillips 1978a; Peterson et al. 1982; Boroson \& Meyers 1992); it is a 
relatively strong Fe II emitter with R$_{4570}$=0.50 (Boroson \& Meyers 1992).
It has however been classified as a Seyfert 1 rather than an NLS1 by Moran et 
al. (1996).

 It is a X-ray source, variable on a time scale of 1\,000 s (Rao et al. 1992). 

 We have not observed this object. \\

 {\bf IRAS~04416+1215} (0441+12) is an NLS1 with an H$\beta$ FWHM =1\,670 km 
s$^{-1}$ (Moran et al. 1996).
 The [O III] lines are very broad (FWHM=1\,150 km s$^{-1}$ according to Moran et
al. 1996); we find 1\,320 km s$^{-1}$. \\

 {\bf IRAS~04576+0912} (0457+09) is an NLS1 according to Moran et al. (1996)
with H$\beta$ FWHM=1\,220 km s$^{-1}$ and R$_{4570}$=1.51. It is a {\it ROSAT} 
X-ray source with a poorly determined photon index $\Gamma$=1.4$\pm$1.0 (Boller 
et al. 1992). 

 The [O III] lines are very broad (FWHM=1\,310 km s$^{-1}$ according to Moran et
al. 1996; we find 1\,290 km s$^{-1}$). \\

 {\bf IRAS~04596$-$2257} (0459$-$22) is an NLS1 with H$\beta$ FWHM $<$ 1\,500 km
s$^{-1}$ (Green et al. 1989). 

 We have not observed this object. \\

 {\bf IRAS~05262+4432} (0526+44) is an NLS1 with H$\beta$ FWHM =700 km s$^{-1}$ 
(Moran et al. 1996).

 The narrow lines are clearly extended on our spectra, with 
$\lambda$6583/H$\alpha$ $\sim$ 0.4 outside the nucleus; the 33 \AA\ mm$^{-1}$
red nuclear spectrum is well fitted by a Lorentzian (695 km s$^{-1}$ FWHM) and 
two sets of narrow components with the same $\lambda $6583/H$\alpha$ ratio 
$\sim$0.4 clearly coming from a H II region. The blue narrow line spectrum is 
composite: a H II region with weak [O III] emission and a Seyfert 2 nebulosity 
with weak H$\beta$ which is too weak to be detected in either H$\alpha$ nor 
[N II]. 

 It is a {\it ROSAT} X-ray source with $\Gamma$=5.0, but the uncertainty in this
value is large because of the high column density (Boller et al. 1992). It is 
located at a low galactic latitude: b=5.7\degr with N$_{\rm H}\sim$ 
3.8$\times$10$^{21}$ cm$^{-2}$. \\

 {\bf RX~J07527+2617} (0749+26) is an NLS1; the Balmer line FWHM is $\sim$
1\,000 km s$^{-1}$ (Bade et al. 1995). \\ 

 {\bf Mark~382} (0752+39) is an NLS1; the H$\beta$ FWHM is 1\,500 (Osterbrock 
1977a) or 1\,400 km s$^{-1}$ (Peterson et al. 1982). We measured 1\,280 km 
s$^{-1}$. \\

 {\bf Mark~110} (0921+52) is a Seyfert 1 (Hutchings \& Craven 1988). The 
H$\beta$ FWHM is in the range 1\,670--2\,500 km s$^{-1}$ (Osterbrock 1977a; 
Crenshaw 1986; Boroson \& Green 1992; Bischoff \& Kollatschny 1999).
 The Fe II emission is weak (R$_{4570}$=0.09--0.16) (Osterbrock 1977a; Meyers \&
Peterson 1985; Boroson \& Green 1992); the Fe II line flux remains constant 
while the Balmer line fluxes vary (Bischoff \& Kollatschny 1999). 

 We have not observed this object. \\

 {\bf Mark~705} (0923+12). The H$\beta$ FWHM is in the range 1\,670--2\,400 
km s$^{-1}$ (Zheng \& O'Brien 1990; Miller et al. 1992; Boroson \& Green 1992).
We found that the broad Balmer components have a FWHM of $\sim$ 1\,770 km 
s$^{-1}$. Coronal lines have been observed by Erkens et al. (1997). \\

 {\bf Mark~707} (0934+01) is an NLS1; the H$\beta$ FWHM is 1\,320 (Boroson \& 
Green 1992) or 2\,000 km s$^{-1}$ (Miller et al. 1992). \\

 {\bf Mark~124} (0945+50) is an NLS1 (De Grijp et al. 1992); the FWHM of the 
broad H$\beta$ component is in the range 1\,050--1\,400 km s$^{-1}$ (Osterbrock 
1977a; Phillips 1978a). Grandi (1978) did not find high excitation lines. \\

 {\bf Mark~1239} (0949$-$01) is an NLS1 (Osterbrock \& Pogge 1985); the FWHM of
the broad Balmer components is $\sim$ 1\,000 km s$^{-1}$ (Osterbrock \& Pogge 
1985; de Robertis \& Osterbrock 1984). In contrast, Rodr\'{\i}guez-Ardila et al.
(2000b) gave 2\,968 and 2\,278 km s$^{-1}$ FWHM for the broad H$\beta$ and 
H$\alpha$ component respectively but called it however an NLS1, while Rafanelli 
\& Bonoli (1984) measured 5\,000 and 4\,800 km s$^{-1}$ respectively. We have 
measured 1\,075 and 905 km s$^{-1}$ respectively, confirming that the object is 
an NLS1.
 Coronal lines have been observed by Rafanelli \& Bonoli (1984) and Erkens et 
al. (1997). \\

 {\bf IRAS~09571+8435} (0957+84) is an NLS1 with H$\beta$ FWHM= 1\,120 km 
s$^{-1}$ and R$_{4570}$=1.46 (Moran et al. 1996).

 Boller et al. (1992) found that the {\it ROSAT} photon index is $\Gamma$=1.39,
a very small value for an NLS1, but the uncertainty is large. \\

 {\bf PG~1011$-$040} (1011$-$04). The H$\beta$ FWHM is 1\,440 (Boroson \& Green
1992) or 1\,980 km s$^{-1}$ (Miller et al. 1992). The spectrum published by 
Simpson et al. (1996) suggests that it is indeed an NLS1.
 
 PG~1011$-$040 has been detected by {\it ASCA} as an X-ray source with a photon
index $\Gamma$=1.93$\pm$0.41; its 2--10 keV luminosity is 6.3$\times$10$^{41}$ 
erg s$^{-1}$; it is a weak X-ray source for its optical luminosity (Gallagher et
al. 2001). \\

 {\bf PG~1016+336} (1016+33) is an NLS1 (Osterbrock \& Pogge 1987). The 
H$\beta$ FWHM is 1\,310 km s$^{-1}$ and R$_{4570}$=0.87 (Goodrich 1989). \\

 {\bf Mark~142} (1022+51) is an NLS1; the H$\beta$ FWHM is in the range
1\,350--1\,790 km s$^{-1}$ (Osterbrock 1977a; Phillips 1978a; Boroson \& Green 
1992; Grupe et al. 1999). \\

 {\bf KUG~1031+398} (1031+39). The broad component of the Balmer lines is 
relatively narrow and, consequently, this object has been classified as an NLS1
by Puchnarewicz et al. (1995).
 The profile of the emission lines in KUG~1031$+$398 is complex; four emission 
components are present: an extended H II region, two distinct Seyfert-type 
clouds identified with the NLR (one of which has quite broad lines: 
$\sim$1\,115 km s$^{-1}$ FWHM), and a relatively narrow ``broad line" component
(1\,060 km s$^{-1}$ FWHM) (Gon\c{c}alves et al. 1999b). 

 According to Pounds et al. (1995), this object is unlike some other 
steep-spectrum (soft X-ray) AGNs in showing a marked absence of rapid X-ray 
variability and of strong Fe II line emission. \\

 {\bf RX~J10407+3300} (1037+33). The FWHM of the Balmer lines is 1\,700 km 
s$^{-1}$ and R$_{4570}$=0.56 (Bade et al. 1995) suggesting that it is an NLS1. \\

 {\bf Mark~734} (1119+12). The spectrum published by Simpson et al. (1996) 
suggests that it is an NLS1; the H$\beta$ FWHM is 1\,820 (Boroson \& Green 1992)
or 1\,940 km s$^{-1}$ (Miller et al. 1992). \\

 {\bf Mark~739E} (1133+21). The double nucleus nature of Mark 739 was first 
described by Petrosian et al. (1979). The eastern component is an NLS1 with 
H$\alpha$ FWHM=900 km s$^{-1}$ and very strong Fe II emission (Netzer et al. 
1987; Mazzarella \& Boroson 1993). Our Lorentzian fit to the Balmer lines has a 
FWHM of $\sim$1\,500 km s$^{-1}$, substantially larger than the published value.
 \\

 {\bf MCG~06.26.012} (1136+34) is an NLS1; the H$\beta$ FWHM is equal to 
1\,685 km s$^{-1}$ (Grupe et al. 1999). Our observations confirm this 
classification. \\

 {\bf Mark~42} (1151+46) is an NLS1 with relatively strong Fe II emission 
(Osterbrock \& Pogge 1985). The H$\beta$ emission line is very narrow, in the 
range 550--670 km s$^{-1}$ FWHM (Osterbrock \& Pogge 1985; Phillips 1978a). 
Grandi (1978) detected [Fe VII]$\lambda$6087.

 On our spectra, the narrow lines are very narrow ($<$ 200 km s$^{-1}$ FWHM 
corrected for the instrumental broadening) and the line ratios are those of a
H II region. \\

 {\bf NGC~4051} (1200+44) is an NLS1 (Leighly 1999a). The H$\beta$ FWHM is 990 
(De Robertis \& Osterbrock 1984) or 1\,150 km s$^{-1}$ (Leighly 1999b). The 
coronal lines are strong (Grandi 1978; Veilleux 1988; Erkens et al. 1997).
 Peterson et al. (2000) found that the Balmer lines could arise in a disk-like
configuration and the high-ionization lines in an outflowing wind, of which we 
observe preferentially the near side. The structure and the kinematics of the
[O III] lines also suggest an outflow (Christopoulou et al. 1997). 

 Our observations show the presence of two components in the [O III] lines, a
narrow one (200 km s$^{-1}$ FWHM) and a broader one (665 km s$^{-1}$ FWHM)
blueshifted by 175 km s$^{-1}$ with respect to the first.

 The X-ray source is variable by at least a factor 30 (Papadakis \& Lawrence 
1995; Leighly 1999a; Uttley et al. 1999; Komossa \& Meerschweinchen 2000); long
term variations in the average X-ray flux might in principle be caused by 
absorption by a varying column of material along the line of sight; but this is 
ruled out by the spectral data; it is the 2--10 keV luminosity which shows large
amplitude long-term variations (Uttley et al. 1999). \\

 {\bf PG~1211+143} (1211+14). The broad Balmer component FWHM has been 
measured to be in the range 
1\,500--1\,860 (Zheng \& O'Brien 1990; Stirpe 1990; 1991; Appenzeller \& 
Wagner 1991; Miller et al. 1992; Boroson \& Green 1992; Wilkes et al. 1999),
except for Miller et al. (1992) who found 2\,280 km s$^{-1}$ for the FWHM of
H$\alpha$. The [Fe VII] line was detected by Appenzeller \& Wagner (1991).

 The soft (0.1--2 keV) X-ray flux varied by at least a factor of 16 (Yaqoob et 
al. 1994). \\

 {\bf Mark~766} (1215+30) is an NLS1 (Osterbrock \& Pogge 1985). The H$\beta$ 
FWHM is 1\,600 (Gonz\'alez Delgado \& P\'erez 1996) or 2\,400 km s$^{-1}$ 
(Osterbrock \& Pogge 1985). Our own measurements give 1\,150 and 1\,630 km 
s$^{-1}$ for the broad H$\alpha$ and H$\beta$ components respectively. The 
spectrum shows relatively strong Fe II emission (Meyers \& Peterson 1985; 
Gonz\'alez Delgado \& P\'erez 1996) and coronal lines (Veilleux 1988; Gonz\'alez
Delgado \& P\'erez 1996). 
 The nucleus shows circumnuclear emission, the spectrum of which is well fitted 
by H II region models (Gonz\'alez Delgado \& P\'erez 1996). \\

 {\bf MS~12170+0700} (1216+07) has been identified with an AGN (Maccacaro et 
al. 1994). Our spectra show strong broad Balmer components with FWHM equal to 
1\,405 and 1\,765 km s$^{-1}$ for H$\alpha$ and H$\beta$ respectively; the 
narrow line system is most probably a H II region as shown by the line ratios 
($\lambda$5007/H$\beta$=1.2 and $\lambda$6583/H$\alpha$=0.60). \\

 {\bf MS~12235+2522} (1223+25) is an NLS1 with H$\beta$ FWHM=1\,730 km 
s$^{-1}$ and $\lambda$5007 FWHM=1\,700 km s$^{-1}$ according to Stephens (1989).
 The broad Balmer components in our spectra were both fitted by a single 
Lorentzian (FWHM$\sim$750 km s$^{-1}$). \\

 {\bf IC~3599} (1235+26). An optical spectrum, taken in May 1991, shows 
permitted lines with widths $\sim$1\,200--1\,500 km s$^{-1}$ showing that this
object is an NLS1; the forbidden lines are narrow and weak (R$_{5007}$ $<$ 0.1)
(Brandt et al. 1995; Mason et al. 1995). Low dispersion spectra taken by Grupe 
et al. (1995; 1999) from 1992 to 1995 show narrow Balmer and forbidden lines 
placing this object close to the borderline between Seyfert 2 and H II galaxies 
in the diagnostic diagrams of Veilleux \& Osterbrock (1987); the resolution used
was unsufficient to clearly show the composite nature of the spectrum; during 
this period, the line ratio R$_{5007}$ was constant (3.3$\pm$0.3). Assuming that
the [O III]$\lambda$5007 line flux is not variable, it follows that, between May
1991 and February 1992, the flux of the Balmer lines has decreased by a factor 
$\sim$27 (Grupe et al. 1995). [Fe VII] $\lambda$6087 have been detected (Komossa
\& Bade 1999).

 Our spectra were taken in March 1997; the line ratio R$_{5007}$  was then 
$\sim$3.3. The H$\beta$ line was too weak for a significant fit to be made, but
the broad H$\alpha$ component could be fitted with a Lorentzian profile with a 
FWHM of $\sim$500 km s$^{-1}$.

 IC~3599 has been detected as a {\it ROSAT} X-ray source (Bade et al. 1995).
The 0.1--2.5 keV X-ray spectrum is extremely steep and is variable by an 
extremely large amount; the count-rate decreased by a factor of $\sim$80 from
December 1990 to June 1992 and then by an additional factor of $\sim$2 to June 
1993 (Grupe et al. 1995; Brandt et al. 1995; Mason et al. 1995). \\

 {\bf PG~1244+026} (1244+02) is an NLS1; the FWHM of the H$\beta$ line is
830 (Boroson \& Green 1992) or 1\,350 km s$^{-1}$ (Miller et al. 1992). \\

 {\bf NGC~4748} (1249$-$13) is an NLS1 with strong Fe II emission (Osterbrock 
\& de Robertis 1985; Moran et al. 1996). The H$\beta$ FWHM is in the range 
1\,100--1\,500 km s$^{-1}$ (Osterbrock \& de Robertis 1985; Maza \& Ruiz 1989; 
Winkler 1992). Rodr\'{\i}guez-Ardila et al. (2000b) measured $\sim$2\,350 km 
s$^{-1}$ for the FWHM of the broad Balmer components. Our own measurements show
that they have a FWHM of $\sim$1\,500 km s$^{-1}$. \\

 {\bf Mark~783} (1300+16) is an NLS1, however the Fe II emission is very weak: 
R$_{4570}<$0.11 (Osterbrock \& Pogge 1985). \\

 {\bf R~14.01} (1338$-$14) is an NLS1 with H$\beta$ FWHM=1\,790 km s$^{-1}$ 
(Maza \& Ruiz 1989). \\

 {\bf Mark~69} (1343+29). Osterbrock (1977a) noted that it is a Seyfert 1 with 
relatively narrow ``broad" Balmer lines ($\sim$1\,500 km s$^{-1}$ FWHM). The 
emission lines on our spectra are well fitted with a single Lorentzian for the 
broad Balmer component and one set of Gaussians for the narrow lines, with line
ratios indicating a H II region ($\lambda$5007/H$\beta$=2.23 and 
$\lambda$6583/H$\alpha$=0.67). \\

  {\bf 2E~1346+2646} (1346+26) is a Seyfert 1 with a relatively narrow ``broad"
H$\beta$ component (Hill \& Oegerle 1993). Our red spectrum is equally well 
fitted with either a broad Gaussian (FWHM=1680 km s$^{-1}$) or a broad 
Lorentzian (FWHM=1235 km s$^{-1}$). On the blue spectrum, the broad H$\beta$ 
component is weak and the measurement quite uncertain. \\
 
 {\bf PG~1404+226} (1404+22) is an NLS1 with a narrow H$\beta$ line:
880 (Boroson \& Green 1992) or 1\,290 km s$^{-1}$ FWHM (Miller et al. 1992). 

 The X-ray flux has changed by a factor 13.1 in 10 years (Forster \& Halpern 
1996) and by a factor 4 in $\sim$3$\times$10$^{4}$ s (Ulrich et al. 1999). \\

 {\bf Mark~684} (1428+28) is an NLS1 (Osterbrock \& Pogge 1987) with prominent
Fe II emission (Persson 1988).

 On the red spectrum, the broad line component is poorly fitted by a single 
Lorentzian; a second, Gaussian component is needed. The broad H$\beta$ line
is too weak for a meaningful fit to be made. The spectrum is composite in the 
sense that the narrow line region has two components, one with relatively broad
lines, the other with very narrow lines and line ratios typical of a H II 
region. \\

 {\bf Mark~478} (1440+35) is an NLS1 (Gondhalekar et al. 1994; Moran et al. 
1996). The H$\beta$ FWHM is in the range 1\,300--1\,915 km s$^{-1}$ (Phillips 
1978a; Peterson et al. 1982; Boroson \& Green 1992; Gondhalekar et al. 1994; 
Grupe et al. 1999). The Fe II emission is strong (Phillips 1977). Grandi (1978)
could not detect high excitation lines. \\

 {\bf PG~1448+273} (1448+27) is an NLS1; the H$\beta$ FWHM is in the range 
910--1\,200 km s$^{-1}$ (Stirpe 1991; Boroson \& Green 1992). \\

 {\bf IRAS~15091$-$2107} (1509$-$21) is an NLS1 (Osterbrock \& de Robertis 
1985), although Moran et al. (1996) rather classified it as a Seyfert 1. 
Goodrich (1989) measured H$\beta$ FWHM= 1\,480 km s$^{-1}$ and R$_{4570}$= 0.54.
Winkler (1992) and Maza \& Ruiz (1989) measured 2\,000 and 1\,600 km s$^{-1}$ 
respectively for the H$\beta$ FWHM.

 We have not observed this object. \\

 {\bf MS~15198$-$0633} (1519$-$06) is an AGN (Margon et al. 1985). The H$\beta$
FWHM is equal to 1\,304 km s$^{-1}$; the [Fe VII]$\lambda$6087 line was not 
detected (Appenzeller \& Wagner 1991).
 We have only a red spectrum with a relatively poor signal-to-noise ratio. \\

 {\bf Mark~486} (1535+54). The H$\beta$ FWHM is in the range 1\,410--1\,650 km
s$^{-1}$ (Boroson \& Green 1992; de Robertis 1985; Osterbrock \& Shuder 1982; 
Boroson et al. 1985). The Fe II emission is relatively strong (Phillips 1978a).
Erkens et al. (1997) have observed coronal lines.

 The X-ray source has an {\it ASCA} photon index $\Gamma$=2.02 $\pm$0.93 and a 
high neutral absorption column density (N$_{\rm H}$=1.2$\times$ 10$^{23}$ 
cm$^{-2}$ plus an unabsorbed component scattered by electrons towards the 
observer; its 2--10 keV luminosity is 1.3$\times$10$^{42}$ erg s$^{-1}$; if the 
optical nucleus was also absorbed by such a large column density, it would not 
be observable and the galaxy would appear as a Seyfert 2 (Gallagher et al. 
2001). \\

 {\bf IRAS~15462$-$0450} (1546$-$04) has been identified with the northern 16.6
mag spiral member of a loose interacting pair (Strauss et al. 1992; Duc et al. 
1997). It is an ultraluminous IR galaxy with a Seyfert 1 spectrum and strong
Fe II lines (Duc et al. 1997); the [O III] lines are broad with FWHM=1\,560 
km s$^{-1}$ (Kim et al. 1998).

 On our spectra, the broad Balmer components have a FWHM of $\sim$1\,700 km 
s$^{-1}$; the narrow lines have a complex profile and can be fitted with two 
Gaussian systems: one has very narrow lines ($<$180 km s$^{-1}$ FWHM corrected 
for instrumental broadening), with line ratios indicating that it is a H II 
region; the second system has very broad lines ($\sim$ 1\,600 km s$^{-1}$ FWHM) 
and correspond to a Seyfert 2 nebulosity. \\

 {\bf Mark~493} (1557+35) is an NLS1 with H$\beta$ FWHM = 410 km s$^{-1}$ 
(Osterbrock \& Pogge 1985). \\

 {\bf EXO~16524+3930} (1652+39) is an NLS1 with Balmer line FWHMs equal to
1\,000 km s$^{-1}$ (Bassani et al. 1989), which is confirmed by our own 
observations. \\

 {\bf B3~1702+457} (1702+45) is an NLS1 according to Moran et al. (1996) 
and Wisotzki \& Bade (1997) who give 490 and 800 km s$^{-1}$ respectively for 
the H$\beta$ FWHM. Leighly (1999b) measured R$_{4570}$=1.86.

 Komossa \& Bade (1998) have shown the presence of a warm absorber. The 
{\it ASCA} spectrum is well fitted by a single power law 
($\Gamma$=2.20$\pm$0.06) plus Galactic absorption, a warm absorber and no soft 
excess (Leighly 1999b; Vaughan et al. 1999a). \\

 {\bf RX~J17450+4802} (1743+48) is a Seyfert 1 (Perlman et al. 1996). The 
Balmer line FWHM is 1\,600 km s$^{-1}$ (Bade et al. 1995). Our blue spectrum
shows that the H$\beta$ broad component FWHM is 1\,355 km s $^{-1}$ and
R$_{4570}$=0.78, so this object is an NLS1. \\

 {\bf Kaz~163} (1747+68) is the southern member of an interacting pair. It is an
NLS1 (Stephens 1989). The H$\beta$ FWHM is in the range 1\,040--2\,110 km
s$^{-1}$ (Kriss \& Canizares 1982; Stephens 1989; Goodrich 1989; Leighly 
1999b). \\

 {\bf Mark~507} (1748$+$68) has been variously classified as a HII region
(Terlevich et al. 1991), a Seyfert 2 (Koski 1978) or a Liner with a 
``transition" type nucleus (Heckman 1980). It is however an NLS1 according to 
Halpern \& Oke (1987), Goodrich (1989), Moran et al. (1996) and Leighly (1999b).
The H$\beta$ FWHM is 965 (Goodrich 1989) or 1\,150 km s$^{-1}$ (Leighly 1999b). 
The Fe II emission is strong with  R$_{4570}$=2.71 (Goodrich 1989) or 1.45 
(Leighly 1999b).

 Our spectra show narrow ``broad" Balmer components ($\sim$1\,335 km s$^{-1}$ 
FWHM). The [O III] lines and the narrow component of H$\beta$ are best fitted by
two sets of Gaussians, one with narrow components and weak [O III] lines, 
typical of a H II region, the other with much broader lines and a large
$\lambda$5007/H$\beta$ ratio. The presence of the H II region component is not 
very surprising as Halpern \& Oke (1987) have found that the emission lines away
from the nucleus are similar to those of a H II region. 

 The {\it ROSAT} photon index is $\Gamma$=1.68$\pm$0.16 (Leighly 1999b), with an
intrinsic neutral hydrogen column density: N$_{\rm H}$=27$\times$10$^{20}$ 
cm$^{-2}$ in excess of the Galactic value (Iwasawa et al. 1998; Leighly 1999b).
The photon index of this object is very small for an NLS1; this could be due to
the presence of a high column density. \\

 {\bf HS~1817+5342} (1817+53). A spectrum published by Engels et al. (1998) 
suggests that this object could be an NLS1. Our spectra show that the broad 
Balmer line component FWHM are $\sim$1\,620 km s$^{-1}$ and R$_{4570}$=0.59,
so there is no doubt that this is an NLS1. \\

 {\bf HS~1831+5338} (1831+53). A spectrum published by Engels et al. (1998) 
suggests that this object could be an NLS1. Our spectra show that the FWHMs of
the broad Balmer components are $\sim$1\,510 km s$^{-1}$ and R$_{4570}$=0.74,
so there is no doubt that this is an NLS1. \\

 {\bf Mark~896} (2043$-$02) is a Seyfert 1 with relatively narrow Balmer lines
and strong Fe II emission (Osterbrock \& Dahari 1983; Morris \& Ward 1988). The 
H$\beta$ FWHM is 1\,390 (Stirpe 1991) or 1\,330 km s$^{-1}$ (Stirpe 1990). 
However Moran et al. (1996) classified it as a Seyfert 1 rather than an NLS1. 
 Our spectra show quite narrow ``broad" Balmer components ($\sim$1\,100 km 
s$^{-1}$) and R$_{4570}$=0.50, so it is an NLS1. \\

 {\bf MS~22102+1827} (2210+18) has been identified with an AGN (Stocke et al. 
1991; Maccacaro et al. 1994). Our spectra show quite narrow ``broad" Balmer
components ($\sim$ 750 km s$^{-1}$), so this is most probably an NLS1. \\

 {\bf Akn~564} (2240+29) is an NLS1 (Goodrich 1989). The H$\alpha$ and H$\beta$
FWHM lie in the ranges 600--730 and 720--1\,030 km s$^{-1}$ respectively 
(Osterbrock \& Shuder 1982; de Robertis \& Osterbrock 1984; Stirpe 1990; 1991; 
Moran et al. 1996; Comastri et al. 2001). The coronal lines are strong 
(Veilleux 1988; Erkens et al. 1997; Comastri et al. 2001).

 The X-ray flux varies by $\sim$50\% in 1.6 h; the X-ray light-curve shows no 
evidence for energy dependence of the variability within the 0.6--10 keV 
bandpass (Turner et al. 1999a; Vaughan et al. 1999b).

 UV absorption lines have been detected, indicating the presence of a warm 
absorber (Crenshaw et al. 1999). \\

 {\bf HS~2247+1044} (2247+10). A spectrum published by Engels et al. (1998) 
suggested that this object could be an NLS1. This classification is confirmed
by our spectra. \\
 
 {\bf Kaz~320} (2257+24) is an NLS1 according to Zamorano et al. (1992) who
measured 1\,700 and 1\,800 km s$^{-1}$ for the FWHM of the broad component of 
H$\alpha$ and H$\beta$ respectively. Our spectra confirm this classification. \\

\normalsize

\begin{acknowledgements}
We thank C. Boisson for a careful reading of the manuscript.
\end{acknowledgements}

\end{document}